\documentclass[aps,prc,reprint,superscriptaddress,showpacs,nofootinbib]{revtex4-1}

\usepackage{color}
\usepackage{framed}
\usepackage{amsmath}
\usepackage{amssymb}
\usepackage{graphicx}
\usepackage[normalem]{ulem}

\def\del{\partial}

\begin{document}

\title{
Composite and elementary nature of a resonance in the sigma model
} 

\author{Hideko Nagahiro}
\thanks{}
\affiliation{Department of Physics, Nara Women's University, 
Nara 630-8506, Japan}
\affiliation{Research Center for Nuclear Physics (RCNP), Osaka
University, Ibaraki, Osaka 567-0047, Japan}

\author{Atsushi Hosaka}
\affiliation{Research Center for Nuclear Physics (RCNP), Osaka
University, Ibaraki, Osaka 567-0047, Japan}

\date{\today}

\begin{abstract}
We analyze the mixing nature of the low-lying scalar resonance
consisting of the $\pi\pi$ composite and the elementary
particle within the sigma model.  
A method to disentangle the mixing is 
{formulated in the scattering theory with the
 concept of the two-level problem.}
 We {investigate} the composite and
 elementary components of the $\sigma$ meson
{by changing a mixing parameter.}
{We also study the dependence of the results on
model parameters such as the cut-off value and the mass of the
 elementary $\sigma$ meson.}
\end{abstract}


\maketitle

%

\section{Introduction}

Understanding the internal structure of hadrons is an important
{issue} to clarify various properties of hadrons.
The purpose of the present work is to shed light upon the nature of
hadrons whether they are {composed} of quarks and gluons confined in a 
single region or rather develop 
hadronic molecular like structure.  
In this paper, the former is referred to as the {\em elementary
hadron}\footnote{The term ``elementary'' here does not necessarily mean 
``elementary'' in ordinary sense, but rather it is used for a degree of
freedom to discuss complex nature of hadrons together with another
degree of freedom, hadronic composite.}
and the latter to as the  {\em hadronic composite}. 
{Here we assume that the above two configurations are well
distinguished, though in general the distinction is not perfect.}

It has been suggested that some hadronic resonances could {well
develop the structure of hadronic composite.}
Especially, the nature of the $\sigma$ meson, which is now listed in the
table of the Particle Data Group (PDG)~\cite{PhysRevD.86.010001},
{has been} a
longstanding
problem~\cite{PhysRevD.2.1680,PhysRevD.49.5779,PhysRevLett.76.1575,PhysRevD.54.1991,PhysRevLett.78.1603,Oller:1997ti,PhysRevD.59.034005,Oller:1998hw,Hyodo:2010jp}.
There are various 
approaches to describe the lowest lying scalar meson by the $q\bar{q}$
state, the four-quark state, the $\pi\pi$ molecular state, and so on.
In general, if more than one quantum state is allowed 
{for a given set of}
quantum numbers, hadronic resonant states are unavoidably
mixtures of these states.  Therefore, an important
issue is to clarify how these components are mixed in a {physical}
hadron. 

{In this paper we}
investigate the mixing nature of the $\sigma$
meson which is treated as a superposition of the {\em elementary
$\sigma$ meson}, and {\em a $\pi\pi$ composite state} within
the framework of the sigma model.  In our previous study~\cite{Nagahiro:2011jn}, 
we have proposed a method to disentangle the
mixture of hadrons having {the elementary component}
and {that of} hadronic
composite by taking the $a_1(1260)$ axial-vector meson as an
example.  
The method makes use of the concept of the two-level problem, and
can be generally applied to other mixed systems if the interaction is
given between the elementary component and constituents {of}
the composite state.  The $a_1$ meson was a good
example because we have a model~\cite{Bando:1987br} involving the
elementary $a_1$ field as 
well as $\pi$ and $\rho$ fields developing the $\pi\rho$ composite
state~\cite{Roca:2005nm,Nagahiro:2008cv}.   

The sigma model also provides us {with a} platform to study the
mixing nature of resonances.  In the so-called nonlinear sigma
model, the $\sigma$ meson is
introduced as a dynamically generated resonance through the
non-perturbative {hadron} dynamics~\cite{Oller:1997ti}.
In that model Lagrangian, the degree of freedom of the elementary
$\sigma$ fields is freezed out by taking the mass of the elementary
$\sigma$ infinite, and instead, the four-pion contact interaction
becomes attractive
to develop the $\sigma$ state as a $\pi\pi$ composite
resonance.

In contrast, in this article, we keep the mass of the elementary
$\sigma$ field finite and treat it as an independent degree of
freedom.  Because the four-pion interaction is still attractive,
the unitarized amplitude {of this interaction develops a pole for the
dynamically generated resonance, which is another degree of freedom for
the $\sigma$ meson.} 
{In the Bethe-Salpeter equation of the sigma model, the two degrees
of freedom couple, and the resulting solution is expressed as a
superposition of the two.}
In this way, we investigate the mixing nature of the $\sigma$ meson
by applying the method given in Ref.~\cite{Nagahiro:2011jn}.

This article is organized as follows. In Sec.~\ref{sec:formalism} we
give the formulation to obtain the non-perturbative scattering
amplitude based on the sigma model.  In Sec.~\ref{sec:2level}, we
introduce the method to disentangle mixture by means of the two-level
problem.  We show our numerical results and give associated discussions
in Sec.~\ref{sec:results}. Finally, Sec.~\ref{sec:sum} is devoted to
conclude this article.

\section{formalism}
\label{sec:formalism}
\subsection{tree level amplitude}
Our discussion of the $\pi\pi$ scattering is based on the sigma
model.  The Lagrangian~\cite{Donoghue:book} is given by
\begin{multline}
 {\cal L}=\frac{1}{4} {\rm Tr} \left( 
\del_\mu \Sigma \del^\mu \Sigma^\dagger
\right) 
+\frac{\mu^2}{4} {\rm Tr}\left(\Sigma^\dagger\Sigma\right)\\
-\frac{\lambda}{16}\left[{\rm Tr}\left(\Sigma^\dagger\Sigma\right)\right]^2
+a {\rm Tr}\left(\Sigma^\dagger+\Sigma\right)\ .
\label{eq:Lagrangian}
\end{multline}
Here in the standard notation the chiral field is parameterized as
$\Sigma=\sigma+i\vec{\tau}\cdot\vec{\pi}$,  and so the first term of
(\ref{eq:Lagrangian}) gives the properly normalized kinetic terms for
$\sigma$ and $\pi$,
\begin{equation}
 \frac{1}{2}(\del_\mu\sigma \del^\mu\sigma +
  \del_\mu\vec{\pi}\cdot\del^\mu\vec{\pi}) \ .
\end{equation}
The second and third terms of {(\ref{eq:Lagrangian})} give the mass and
the four-point interaction
terms of $\sigma$ and $\pi$, {respectively}, where $\mu$ is their common mass and
$\lambda$ the coupling {constant}.  When chiral symmetry is
spontaneously broken, the potential, the sum of the second and third
terms, takes the minimum at a finite expectation value of
${\sigma=}\langle\sigma\rangle\equiv f_\pi$, where $f_\pi$ is the pion decay
constant.   The physical $\sigma$ field is then expanded around this
vacuum, such that $\sigma\rightarrow f_\pi+\sigma$.  
The last term of (\ref{eq:Lagrangian}) is for explicit breaking of
chiral symmetry and gives the physical mass for the pion after the
spontaneous chiral symmetry breaking.

{In this study}, we employ the nonlinear representation for the chiral
field as
\begin{equation}
 \Sigma=(f_\pi+\sigma)U, \ \
  U=\exp\left(\frac{i\vec{\tau}\cdot\vec{\pi}}{f_\pi}\right).
\label{eq:L2}
\end{equation}
This provides interaction terms for $\pi$ and $\sigma$ including three
and four-point interactions as,
\begin{multline}
\label{eq:Lint}
{\cal L}_{\rm int} = 
\frac{1}{6f_\pi^2}\left\{
\left(\del_\mu\vec{\pi}\cdot\vec{\pi}\right)^2
-(\del_\mu \vec{\pi}\cdot\del^\mu\vec{\pi})(\vec{\pi}\cdot\vec{\pi})
\right\} \\
+\frac{m_\pi^2}{24f_\pi^2}(\vec{\pi}\cdot\vec{\pi})^2
+\frac{1}{f_\pi}  \sigma\, \del_\mu \vec{\pi}\cdot
\del^\mu\vec{\pi}
-\frac{m_\pi^2}{2f_\pi} \sigma \vec{\pi}\cdot\vec{\pi}.
\end{multline}
{The three parameters, $\mu$, $\lambda$, and $a$, in the Lagrangian (\ref{eq:Lagrangian}) are determined
by the three inputs, $f_\pi=92.4$~MeV, $m_\pi=138$~MeV
(isospin-averaged) and $m_0$.    The mass of the elementary
$\sigma$ field, $m_0$, is varied in the present study, but we shall start with
the value $m_0=550$~MeV.}

\begin{figure}[hbt]
\includegraphics[width=0.6\linewidth]{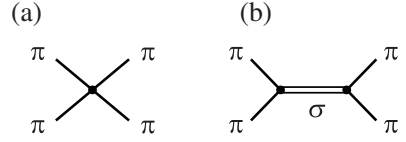}
\caption{Feynman diagrams contributing for the function $A(s)$ in
 Eq.~(\ref{eq:A}).
\label{fig:tree}
}
\end{figure}

The $\pi\pi$ scattering amplitude {at the tree level is} determined
in terms of a single function $A(s,t,u)$ by
\begin{multline}
 T_{\alpha\beta,\gamma\delta}^{\rm tree} =
  A(s,t,u)\delta_{\alpha\beta}\delta_{\gamma\delta}
+ A(t,s,u)\delta_{\alpha\gamma}\delta_{\beta\delta}\\
+ A(u,t,s)\delta_{\alpha\delta}\delta_{\beta\gamma}\ ,
\end{multline}
where $\alpha$, $\beta$, $\gamma$, $\delta$ denote the isospin
components of the pions.
The amplitude with isospin $I=0$ for the $\sigma$ channel is given by
\begin{equation}
 T^{\rm tree}_{I=0} = 3 A(s,t,u) + A(t,s,u) + A(u,t,s)\ .
\label{eq:T0}
\end{equation}
The function $A(s,t,u)(\equiv A(s))$ is obtained from the interaction Lagrangian
(\ref{eq:Lint}) as
\begin{equation}
 A(s) = - \frac{1}{f_\pi^2}(s-m_\pi^2)
+
\frac{1}{f_\pi^2}(s-m_\pi^2)^2\frac{1}{s-m_0^2} \ ,
\label{eq:A}
\end{equation}
where the first term comes from the four-pion contact interaction as
depicted in Fig.~\ref{fig:tree}(a) and the second term the elementary
$\sigma$-exchange in $s$-channel (we refer to it as ``$\sigma$-pole''
hereafter) as in Fig.~\ref{fig:tree}(b).  

{The $s$-wave amplitude for the $\sigma$ meson can be projected
out,} 
\begin{equation}
 v(s) = \frac{1}{2} \int^1_{-1} dx\
  T^{\rm tree}_{I=0}(s,t(x),u(x))P_{\ell=0}(x)\ ,
\end{equation}
{in the center-of-mass frame.}
The result of the projection is given by
\begin{multline}
 v(s) = -\frac{1}{f_\pi^2}(2s-m_\pi^2) 
+\frac{3}{f_\pi^2}(s-m_\pi^2)^2\frac{1}{s-m_0^2}\\
-
\frac{1}{f_\pi^2}
\left[(s-2m_0^2)-\frac{2(m_\pi^2-m_0^2)^2}{s-4m_\pi^2}
\ln\left(
\frac{m_0^2}{m_0^2-4m_\pi^2+s}
\right)
\right]\ ,
\label{eq:t_tree}
\end{multline}
where we {have} use{d} the relation $s + t + u = 4m_\pi^2$ for on-shell
amplitudes.  Here the first term in Eq.~(\ref{eq:t_tree})
is obtained by the four-pion contact interaction,
the second term the $\sigma$-pole, and the last term the
$\sigma$-exchange in $t$- and $u$-channels.

\subsection{Unitarized scattering amplitude}

\begin{figure}[hbt]
\includegraphics[width=0.45\textwidth]{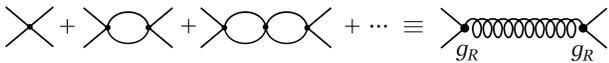}
\caption{Sum of the infinite set of diagrams that contributes to the meson-meson
 scattering amplitude (\ref{eq:BSeq}).   
{If a pole is developed, then it is interpreted as}
the composite state in Eq.~(\ref{eq:t_WT})
{which is depicted as in the right-hand-side.}
\label{fig:WT_pole}
}
\end{figure}

The tree level amplitude projected on the $s$-wave, $v(s)$, is now used
as a potential (interaction kernel) in the Bethe-Salpeter (BS) equation
to find the $\sigma$ meson as a resonance of $\pi\pi$ scattering.  In
literatures, the process is often referred {to} as unitarization.  The
resulting unitarized amplitude $t$ is then given by
\begin{eqnarray}
 t&=&v+vGt \nonumber \\
&=&v+vGv+vGvGv+\cdots\ ,
\label{eq:BSeq}
\end{eqnarray} 
where infinite set of diagrams are summed up as depicted in
Fig.~\ref{fig:WT_pole}.
{In general this is an integral equation, but in many recent
applications, the on-shell factorization is employed to reduce it to an
algebraic equation.  Thus the resulting amplitude $t$ is obtained as
}
\begin{equation}
 t=\frac{1}{v^{-1}-G}\ ,
\label{eq:t}
\end{equation}
where $G$ denotes the integrated $\pi\pi$ two-body propagator {as}
\begin{equation}
 G(\sqrt{s})=\frac{i}{2}\int
  \frac{d^4k}{(2\pi)^4}\frac{1}{(P-k)^2-m_\pi^2+i\epsilon}
\frac{1}{k^2-m_\pi^2+i\epsilon}.
\label{eq:G}
\end{equation}
{Here} $P$ is the total momentum {in the center-of-mass frame,}
 $P=(\sqrt{s},0,0,0)$, and  
 the factor 1/2 is introduced as the symmetry factor for the 
identical particles.  We {evaluate} the regularized loop function in
 Eq.~(\ref{eq:G}) by the dimensional regularization scheme as,
\begin{eqnarray}
 G(\sqrt{s})&=&\frac{1}{32\pi^2}\left[a(\mu) + \ln\frac{m_\pi^2}{\mu^2}
\right.\nonumber\\
&&+\left.
\frac{2q}{\sqrt{s}}\left\{\ln(s+2q\sqrt{s})-\ln(s-2q\sqrt{s}) - \pi i\right\}
\right],\nonumber\\
\label{eq:G_dim}
\end{eqnarray}
where $\mu$ is the {renormalization} scale.
$a(\mu)$ {is} the subtraction
constant at the scale $\mu$,
{and t}he three-momentum {$q$} of two pions in the center-of-mass frame
for a given $s$ is obtained by,
\begin{equation}
 q=\frac{1}{2\sqrt{s}}\lambda^{1/2}(s,m_\pi^2,m_\pi^2)\ ,
\label{eq:q}
\end{equation}
with $\lambda(x,y,z)=x^2+y^2+z^2-2xy-2yz-2zx$.  
We first use the natural value for
$a(\mu)$~\cite{Hyodo:2008xr,Hyodo:2010jp}. 
Later, we employ another form regularized by the three-dimensional
cut-off as
\begin{equation}
 G(\sqrt{s})=\frac{1}{4\pi^2}\int_0^\Lambda  \frac{k^2dk}{E_\pi
(s-4E_\pi^2+i\epsilon)} \ ,
\label{eq:G_cut}
\end{equation}
where $E_\pi=\sqrt{k^2+m_\pi^2}$,
 to investigate 
{cut-off dependence of various properties of the $\sigma$ meson.}

If the potential $v$ is sufficiently attractive,
 the amplitude $t$ in Eq.~(\ref{eq:t}) develops a pole corresponding to a
 bound or resonant state at the
 energy satisfying $v^{-1}-G=0$.  
When we need to find a bound state pole below the threshold, we use the
loop function in the first Riemann sheet ($G^I$) with Im\,$q>0$. 
In the present 
study in the nonlinear representation, the four-pion
 interaction is attractive and the pole appears above the two
 pion threshold in the second Riemann sheet as a resonant state, which 
 can be interpreted as the $\pi\pi$ composite $\sigma$
 meson~\cite{Oller:1997ti}.   
The loop function in the second Riemann sheet ($G^{I\hspace{-.15em}I}$) can be obtained
by 
\begin{equation}
 G^{I\hspace{-.15em}I}(\sqrt{s})=G^{I}(\sqrt{s}) + i \frac{q}{8\pi\sqrt{s}}
\label{eq:GII}
\end{equation}
with Im\,$q>0$~\cite{Roca:2005nm}.
Here, we recall that the generation of the composite $\sigma$ 
 state through the non-perturbative dynamics is a feature of the
 nonlinear representation of the sigma model.   

In this study, we split the tree-level amplitude in
Eq.~(\ref{eq:t_tree}) into two parts, 
the ``contact'' and ``$\sigma$-pole'' terms as,
\begin{eqnarray}
v &=& v_{\rm con} + v_{\rm pole} \label{eq:v}\\
v_{\rm con} &=& -\frac{1}{f_\pi^2}
\left[
3s-m_\pi^2-2m_0^2\right.\nonumber\\
&&-\left.\frac{2(m_\pi^2-m_0^2)^2}{s-4m_\pi^2}
\ln\left(
\frac{m_0^2}{m_0^2-4m_\pi^2+s}
		  \right)\right] 
\label{eq:t_con}
\\
v_{\rm pole} &=& \frac{3}{f_\pi^2}(s-m_\pi^2)^2\frac{1}{s-m_0^2}
\ .
\label{eq:t_pole}
\end{eqnarray}
The ``contact'' interaction $v_{\rm con}$
contains not only the four-pion interaction but also the
contribution of the $\sigma$-exchange in $t$- and 
$u$-channels.  The latter slightly modifies the attractive potential coming
from the four-pion interaction, but the total attraction of $v_{\rm con}$ is
still strong enough {such that the unitarized amplitude $t_{\rm composite}$
\begin{equation}
 t_{\rm composite}=\frac{v_{\rm con}}{1-v_{\rm con}G}
\label{eq:t_com}
\end{equation}
develops a resonance pole.}
{In the present work, we identify the $\sigma$ meson generated 
in Eq.~(\ref{eq:t_com}) with the {\em composite} $\sigma$ meson.}
%
{In contrast, $v_{\rm pole}$ has the {\em elementary} $\sigma$-pole.  In
this way, we have defined two ``seeds'' of the 
$\sigma$ meson having different origins. 
}

{A similar decomposition of the amplitude into the contact and pole terms
was also considered}
in Ref.~\cite{Hyodo:2010jp}.
{There}
the contact interaction 
$T_{\rm tree}^{\rm (contact)}$ {was} {chosen to be} repulsive, while 
the $v_{\rm con}$ in Eq.~(\ref{eq:t_con}) is attractive.  
This difference comes from the definition of the ``$\sigma$-pole term'',
 namely, 
the authors in Ref.~\cite{Hyodo:2010jp} isolate the
$\sigma$-pole in the {\em linear} representation of the sigma model,
 while we isolate {it in} the {\em nonlinear} representation.  
In the linear representation, the
 $\sigma\pi\pi$ coupling does not depend on the energy, while in the
 nonlinear representation the $\sigma\pi\pi$ coupling does depend on the
 energy.  Furthermore the four-pion interaction in the linear
{representation} is repulsive, and hence one does not have a composite
 $\sigma$-pole.
 
Substituting Eq.~(\ref{eq:v}) as the potential
into the BS equation we obtain the full scattering amplitude in $s$-wave
as
\begin{equation}
 t = \frac{v_{\rm con}+v_{\rm pole}}{1-(v_{\rm con}+v_{\rm
  pole})G}.
\label{eq:T_full}
\end{equation}
{In the present work, we analyze this full scattering amplitude in
detail to study the nature of the $\sigma$ meson.}

\section{reduction to the two-level problem}
\label{sec:2level}
{The two bases of composite and elementary natures mix in the
solution of the full amplitude (\ref{eq:T_full}).  The situation is
similar to a two-level problem in the quantum mechanics.  This idea has
been employed in Ref.~\cite{Nagahiro:2011jn}, where the mixing nature of
$a_1(1260)$ axial vector meson consisting of $\pi\rho$ composite and
elementary $a_1$ meson has been investigated.  There two physical $a_1$
poles are found as 
superpositions of the two basis states.  One of them is identified {with}
the experimentally observed $a_1$ meson, which is found to have
comparable amount of the elementary $a_1$ component to that of the
$\pi\rho$ composite.  }

We apply this method to the
study of the $\sigma$ meson.  
Here we show
details of the formulation which ha{ve}
been omitted in Ref.~\cite{Nagahiro:2011jn}.
We first express the
amplitude $t_{\rm composite}$ {by an $s$-channel}
pole term as
\begin{equation}
   t_{\rm composite}
  \equiv g_R(s)\frac{1}{s-s_p}g_R(s) \ ,
  \label{eq:t_WT}
\end{equation}
where $s_p$ is the pole position of the amplitude {$t_{\rm composite}$}.  In
this form, we can interpret $(s-s_p)^{-1}$ as the one-particle
propagator of the composite $\sigma$ meson.
{Furthermore,}
  $g_R(s)$
defined by Eq.~(\ref{eq:t_WT}) is interpreted as the vertex function
of the composite $\sigma$ meson to two pions in the neighborhood of the
pole, $s\sim s_p$. 
{However, as $s$ is getting apart from $s_p$ this interpretation is
no longer appropriate, where instead it should be reinterpreted together
with background contribution.
}

Having the form of Eq.~(\ref{eq:t_WT}), we can express the full scattering
amplitude (\ref{eq:T_full}) by
\begin{equation}
 t=(g_R,g)
\frac{1}{\hat{D}_0^{-1}-\hat{\Sigma}}
\begin{pmatrix}
 g_R  \\ g
\end{pmatrix} \ ,
\label{eq:T_fullM}
\end{equation}
where 
\begin{equation}
 \hat{D}_0^{-1} = 
\begin{pmatrix} s-s_p & 0  \\ 0  & s-m_0^2 \end{pmatrix},\ 
 \hat{\Sigma} = 
\begin{pmatrix}
0 & g_R G g \\ 
g G g_R & g G g
\end{pmatrix} \ .
\label{eq:D0&Sigma}
\end{equation}
Here $g$ is the coupling of $\sigma\pi\pi$ as $g^2=3(s-m_\pi^2)^2/f_\pi^2$.
The detailed derivation is given in Appendix~\ref{sec:Appendix-A}.
The diagonal elements of $\hat{D}_0$ are the free propagators of two
different $\sigma$'s, one for the composite $\sigma$ and the other for
the elementary $\sigma$.  The matrix 
$\hat{\Sigma}$ expresses the 
self-energy and mixing interaction between the two $\sigma$'s.

\begin{figure}[h]
\includegraphics[width=0.45\textwidth]{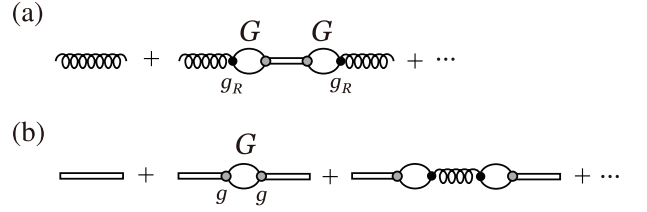}
\caption{
{Full propagators of (a) composite and (b) elementary $\sigma$ mesons
 defined in Eq.~(\ref{eq:full_D}).  The solid line
 indicates the $\pi$ propagator, while the curly and double
 lines are those of the composite and elementary $\sigma$ mesons.} 
\label{fig:full_D}}
\end{figure}

Now, the matrix
\begin{equation}
 \hat{D} = \frac{1}{\hat{D}_0^{-1}-\hat{\Sigma}}
\label{eq:full_D}
\end{equation}
is the full propagators of the physical states
represented by the two bases of the composite and elementary $\sigma$
mesons.  
The diagonal $D^{ii}$ correspond to full propagators
of the composite and elementary $\sigma$ mesons as shown in
Fig.~\ref{fig:full_D}, which express {each} $\sigma$ meson acquires the
quantum effects {through the mixing from the other as well as the
self-energy.} 
{In this manner we can study the mixing nature of the $\sigma$ meson
by analyzing the properties of $\hat{D}$.
}

The important feature of the propagators in Eq.~(\ref{eq:full_D}) is
that they have poles $m^*$ exactly at the same positions as the full
amplitude $t$ in Eq.~(\ref{eq:T_full}).  The residues of the
diagonal elements $D^{ii}$ are obtained by
\begin{equation}
 z^{ii}=\frac{1}{2\pi i} \oint_\gamma D^{ii}(s) ds\ ,  \ \ (i=1,2)
\end{equation}
where $\gamma$ is a closed circle around the pole 
$m^*$.  They {are} the wave function renormalizations {for the
basis states} {$i$} and then
{they have} 
information on the mixing rate of the physical resonant pole.  For
instance, $D^{11}$ is the full propagator of the composite $\sigma$
meson, and its residue $z^{11}$ has the meaning of the probability of
finding the composite $\sigma$ component in the resulting state.   So we
schematically express the 
physical $\sigma$ state $|\sigma\rangle_{\rm phys}$ as
\begin{equation}
 |\sigma\rangle_{\rm phys} = \sqrt{z^{11}}|1\rangle +
  \sqrt{z^{22}}|2\rangle
\label{eq:state}
\end{equation}
with $|1\rangle$ and $|2\rangle$ {the} composite and elementary
states.

The residue $z^{22}$, the wave function
renormalization factor for the elementary 
$\sigma$, can be computed also by
\begin{equation}
 z^{22} =
  \left(1-\left.\frac{d\Pi(s)}{ds}\right|_{s={m^*}^2}\right)^{-1},
\end{equation}
where $\Pi(s)$ denotes the self-energy for the elementary $\sigma$
given by,
\begin{equation}
 \Pi(s)=3\frac{(s-m_\pi^2)^2}{f_\pi^2}\frac{G(\sqrt{s})}{1-v_{\rm
  con}G(\sqrt{s})} .
\end{equation}
The wave function renormalization factor is often used 
to {study} the ``compositeness'' of the physical state.
The relation between the compositeness condition 
discussed in
Refs.~\cite{Weinberg:1965zz,Lurie:1964,*Lurie:book,PhysRevC.85.015201} and the
$z^{22}$ in this two-level problem will be {investigated}
{elsewhere}~{\cite{nagahiro-hosaka}}.

\begin{figure}[h]
\includegraphics[width=0.3\textwidth]{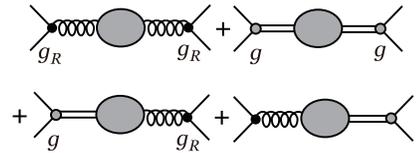}
\caption{Full scattering amplitude in terms of the full propagators
 $\hat{D}$.  The first two diagrams denote the amplitudes through the
 full propagators of composite and elementary $\sigma$ mesons ($D^{11}$
 and $D^{22}$), respectively, and the last two diagrams those
 of off-diagonal propagators ($D^{21}$ and $D^{12}$).
\label{fig:full_T}}
\end{figure}

Now, we can write the full scattering amplitude around {a} pole $m^*$
by using {the components} of the full propagator $\hat{D}$ as,
\begin{eqnarray}
 t &\simeq& g_R\frac{z^{11}}{s-{m^*}^2}g_R
+ g\frac{z^{22}}{s-{m^*}^2}g \nonumber\\
&+&
g\frac{z^{21}}{s-{m^*}^2}g_R+
g_R\frac{z^{12}}{s-{m^*}^2}g 
\nonumber\\
\label{eq:T_fullD}
\end{eqnarray} 
as shown in Fig.~\ref{fig:full_T}.  
{Here, 
the residue of the off-diagonal propagator, $z^{12}(=z^{21})$,
has {the} following relation with $z^{11}$ and $z^{22}$ as
\begin{equation}
 (z^{12})^2=z^{11}z^{22}\ .
\end{equation}
}
Therefore the full amplitude near the pole can be expressed 
by\footnote{The phase ambiguity in taking the square-root of $z^{ii}$ in
Eq.~(\ref{eq:T_full3}) can be absorbed in {the} definition of $g_R$ in
Eq.~(\ref{eq:t_WT}).} 
\begin{equation}
 t=(g_R\sqrt{z^{11}}+g\sqrt{z^{22}})^2\frac{1}{s-{m^*}^2}.
\label{eq:T_full3}
\end{equation}
In a simple Yukawa theory where only one ``seed'' state exists, 
the scattering amplitude is given by
\begin{equation}
 T_{\rm Yukawa}=g_0^2Z\frac{1}{s-{M^*}^2}\ ,
\label{eq:Yukawa}
\end{equation}
where $g_0$ is the original Yukawa coupling constant and $Z^{1/2}$ the
wave function renormalization.  Comparing Eqs.~(\ref{eq:T_full3}) and
(\ref{eq:Yukawa}), we realize that the renormalized coupling constant
$g_0Z^{1/2}$ is now replaced by the sum of those of the two bases
$g_R\sqrt{z^{11}}$ and $g\sqrt{z^{22}}$.  
In this form of Eq.~(\ref{eq:T_full3}), we can clearly see that the
contributions from the original basis states to the full scattering
amplitude is determined by probabilities of finding
their components in the physical state multiplied by
their couplings to the scattering state.

\section{Numerical Results}
\label{sec:results}
\subsection{pole-flow in complex-energy plane}

\begin{table*}[htb]
\begin{center}
\begin{tabular}{c|c|l} 
parameter & pole position (in unit of MeV)&  property of pole (mark in
 Figs.~\ref{fig:pole_flow_hyodo} and/or \ref{fig:pole_flow})
 \\\hline\hline
$x=0$ & $360.8-354.6i$  &  naive-composite $\sigma$ generated by
	 four-pion interaction ($\blacksquare$)\\
      & {$550 -0i$}      &  {elementary $\sigma$ (${\circ}$)}    \\
$x=1$ & $420.5-128.0i$  &  physical state (\textbullet)\\\hline
$X=0$ & $390.7-308.4i$  &  composite $\sigma$ generated by $v_{\rm con}$
	 ($\blacktriangle$)\\ 
      & {$550-0i$}      &  {elementary $\sigma$ (${\circ}$)}    \\
$X=1$ & $420.5-128.0i$  &  physical state (\textbullet) \\\hline\hline
\end{tabular}
\end{center}
\caption{Pole positions of {the full propagator $\hat{D}$} {when} the mass
 of the elementary $\sigma$ meson $m_0=550$~MeV {is employed.}}
\label{tab:pole_position}
\end{table*}

\subsubsection{Parameterization-(I)}
\begin{figure}[t]
\includegraphics[width=0.95\linewidth]{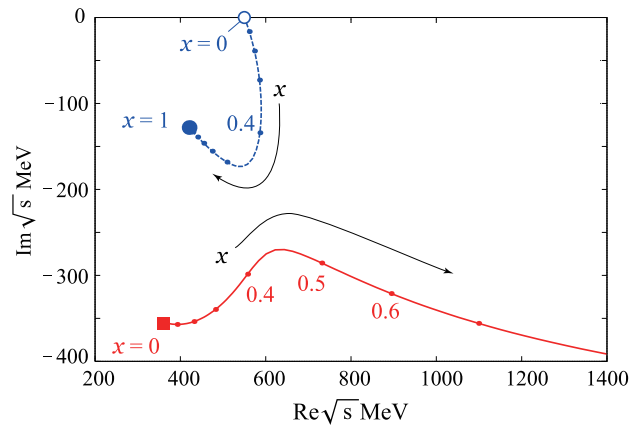}
\caption{(color online) Trajectories of the poles of $\hat{D}$
 by changing the parameter $x$.  The open circle ($\circ$) indicates the
 mass of the elementary $\sigma$ meson, $m_0$, and the solid square
 ($\blacksquare$) the
 pole position of the composite $\sigma$ generated by the four-pion
 interaction only.
 The solid circle (\textbullet) indicates the pole position of the physical state
 at $x=1$.
\label{fig:pole_flow_hyodo}}
\end{figure}

Now, the full scattering amplitude in Eq.~(\ref{eq:T_full}) or
Eq.~(\ref{eq:T_fullM}) {and hence the full propagator} can be {calculated}
numerically. {W}e {first}
investigate the pole position  
of $\hat{D}$ in complex-energy plane {by} vary{ing} the coupling
strength of the $\sigma\pi\pi$ three-point vertex.
{To this end, w}e introduce the parameter $x$ as,
\begin{equation}
 A(s,t,u) = -\frac{1}{f_\pi^2}(s-m_\pi^2) + x
  \frac{(s-m_\pi^2)^2}{f_\pi^2}\frac{1}{s-m_0^2}  
\label{eq:A2}
\end{equation} 
and the same for $A(t,s,u)$ and $A(u,t,s)$ { as in
Ref.~\cite{Hyodo:2010jp}}.  
We call this ``parameterization-(I)''. 
In Fig.~\ref{fig:pole_flow_hyodo}, we show the resulting pole-flow in
complex-energy plane by changing the parameter $x$ from 0 to 1.  Here we take
the mass of the elementary $\sigma$ meson $m_0=550$~MeV.

At $x=0$, $\hat{D}$ has two poles at $\sqrt{s}=550$~MeV and
$\sqrt{s}=360.8-354.6i$~MeV (open circle and solid square
in
Fig.~\ref{fig:pole_flow_hyodo}).  The former pole does
not appear  
in the $\pi\pi$ scattering amplitude {trivially because $x=0$}.
The latter corresponds purely to 
the $\pi\pi$ composite $\sigma$ dynamically generated by the four-pion
interaction. 
We summarize {the pole positions and their properties} in
Table~\ref{tab:pole_position}.

{For finite values of $x \lesssim 0.45$, the two poles come closer
to each other.  After that, ``level-crossing \&
width-repulsion''~\cite{Nawa:2011pz} takes place, {and} the pole
starting from the  elementary $\sigma$ ends at
$\sqrt{s}=420.5-128.0i$~MeV {(solid circle in
Fig.~\ref{fig:pole_flow_hyodo})}.  The other pole 
from the composite $\sigma$ moves to higher energy region rapidly and,
interestingly, it disappears exactly at $x=1$.  }
We come back to this point later.

\begin{figure}[htbp]
\includegraphics[width=0.95\linewidth]{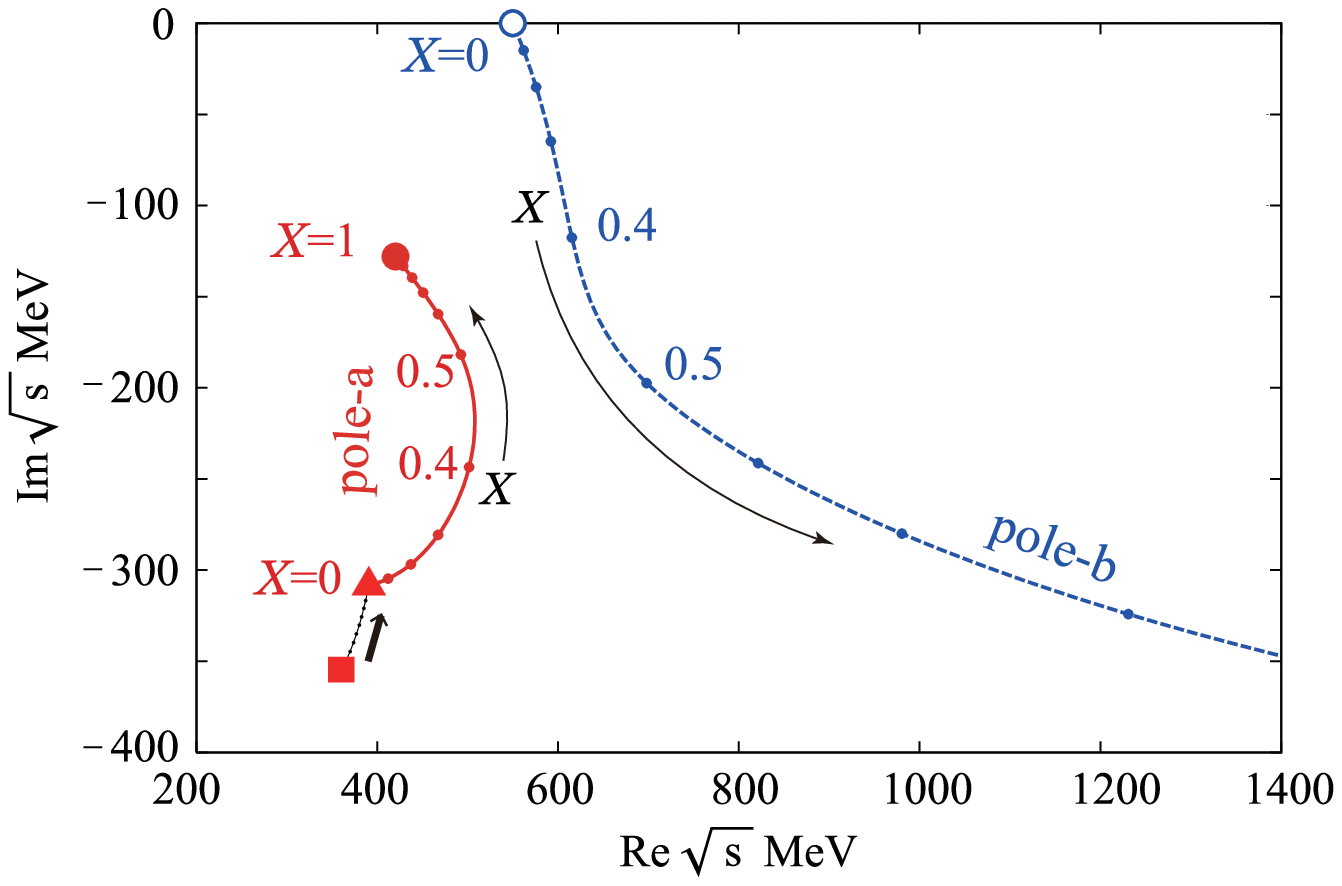}
\caption{(color online) Trajectories of the poles of $\hat{D}$ by
 changing the mixing parameter $X$.  The open circle ($\circ$) indicates
 the mass of the elementary $\sigma$ meson, $m_0$, the solid square
 ($\blacksquare$) the pole position 
 of the composite $\sigma$ generated by the four-pion interaction only.
 The solid triangle ($\blacktriangle$) indicates the pole position of the composite
 $\sigma$ meson developed by the four-pion interaction and the
 $\sigma$-exchange in $t$- and $u$-channels.    The solid circle
 (\textbullet) indicates the
 pole position of the physical state at $X=1$.   The thick arrow shows the
 shift of the composite pole position by adding the $\sigma$-exchange
 term as explained in text.
\label{fig:pole_flow}}
\end{figure}

\subsubsection{Parameterization-(I\hspace{-.15em}I)}
Next, to investigate the mixing properties {in} the
two-level problem, we introduce the mixing parameter $X$ in front of the
$v_{\rm pole}$ as, 
\begin{equation}
 v(s;X) = v_{\rm con}(s) + X v_{\rm pole}(s)
\label{eq:X}
\end{equation}
which controls the mixing strength {of the elementary $\sigma$ meson
to the amplitude.  
We call this ``parameterization-(I\hspace{-.1em}I)''.
Unlike the case (I),
the $\sigma$-exchange in $t$- and $u$-channels is already included in
$v_{\rm con}$}. 
At $X=0$, we find a pole at 
$\sqrt{s}=390.7-308.4i$~MeV (solid triangle
in Fig.~\ref{fig:pole_flow}), which we have called the composite
$\sigma$-pole in this article. In contrast to this pole, we
refer to the pole generated by the four-pion interaction only (solid square in
Fig.~\ref{fig:pole_flow}) as {\em naive}-composite $\sigma$ to
distinguish them.
  We find that the $\sigma$-exchange
contribution shifts the composite pole
position (as shown by the thick arrow in the figure), but its contribution is small.
When the mixing is turned on, unlike the case (I),
``level-repulsion \& 
width-crossing'' takes place.
The pole starting from the composite
$\sigma$ moves closer to the real axis (we refer to it as ``pole-a'')
and becomes the physical state at $X=1$,
while that from the elementary 
$\sigma$ (``pole-b'') goes far away from the real axis and finally
disappear exactly at
$X=1$.  

{From the above analysis, we observe that the pole-flow is not a
unique one but depends very much on the choice of the flow parameter,
$x$ or $X$.  This further indicates that the nature of the physical pole
at $x$ or $X=1$ does not reflect that of the pole at the original
point, $x$ or $X=0$, connected by the flow}.  

\begin{figure*}[htbp]
\includegraphics[width=0.8\linewidth]{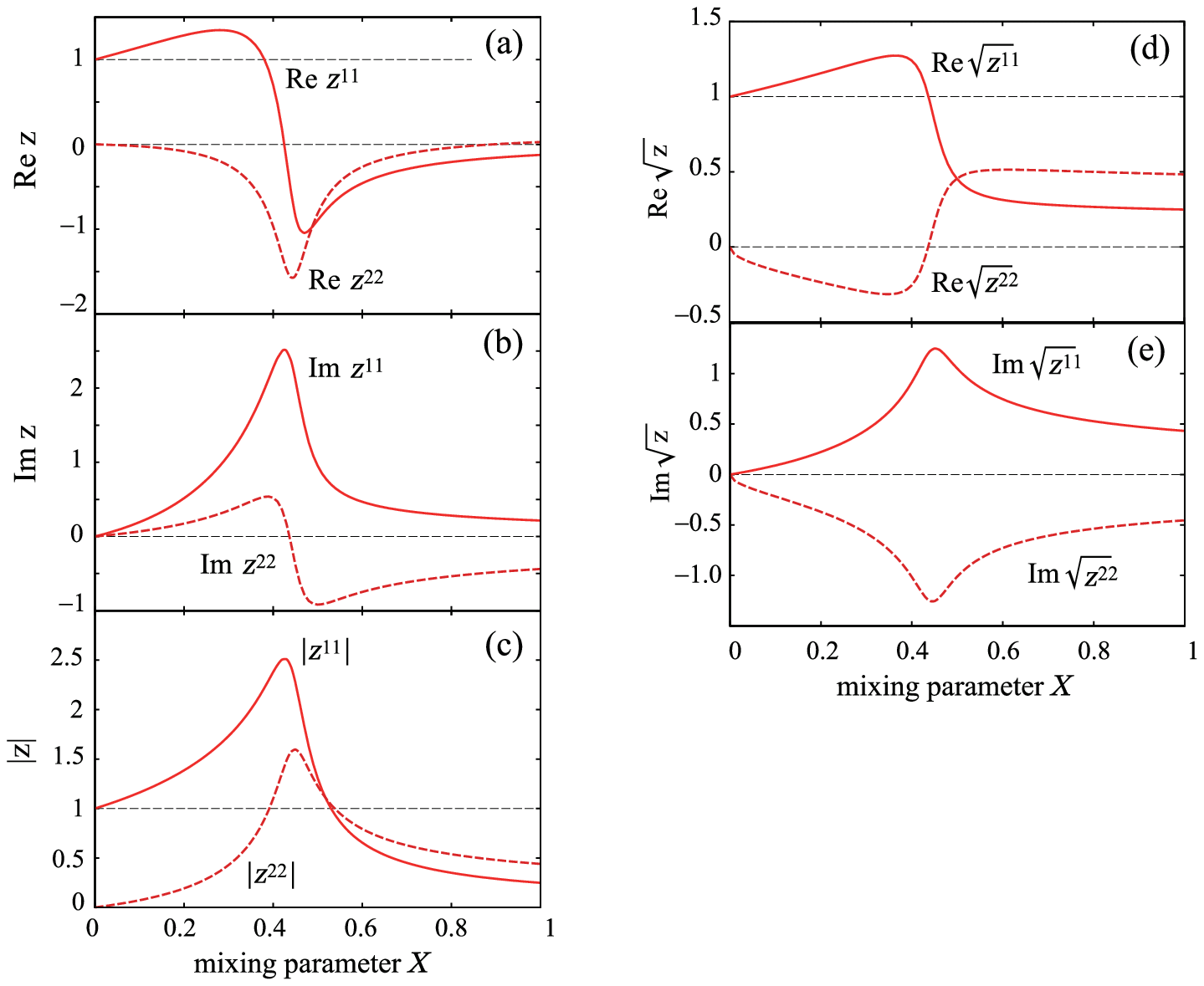}
\caption{(color online) Residues $z^{11}$ and
 $z^{22}$ of the full propagator $D^{11}$ and $D^{22}$ at the pole-a
 shown in
 Fig.~\ref{fig:pole_flow} as functions of the 
 mixing parameter $X$.  
(a) The real parts of the residues, (b) those of the imaginary parts, 
 (c) the modulus of the residues, (d) the real parts of the square-root
 of the residues and (e) those of the imaginary parts are shown, respectively.
\label{fig:allz_vs_X}}
\end{figure*}

\subsection{residues and the nature of the resonance}

{To study the mixing nature of the poles,}
we evaluate the residues $z^{11}$ and $z^{22}$ of pole-a (red solid line
in Fig.~\ref{fig:pole_flow}) with 
parameterization-(I\hspace{-.1em}I). 
In Fig.~\ref{fig:allz_vs_X} 
we show the residues as functions of the mixing parameter $X$.
First, {recall} that at $X=0$ the
pole-a is purely the composite $\sigma$ meson ($z^{11}=1$ and
$z^{22}=0$) as we expected.  

Around $X\sim 0.4$ -- $0.5$, $|z^{11}|$ and $|z^{22}|$ take the
maximum value larger than 1, where  in complex-energy plane pole-a
and pole-b come closest to each other.  This behavior of the residues looks
like ``resonance-shape''.  For example, the line shape of the real
(imaginary) part of $z^{11}$ is similar to the resonance shape of the
real (imaginary) part of a scattering amplitude.  In fact, we can
understand this behavior by explicitly writing down the propagator, for
example, $D^{22}$. {Let} $m_a^*(X)$ and $m_b^*(X)$ {be}
the two pole {values} of the full propagator {as functions of}
$X$, {then} we have
\begin{eqnarray}
 D^{22}(s;X)=\frac{\zeta(s)(s-s_p)}{(s-m_a^{*2}(X))(s-m_b^{*2}(X))}\ ,\\
 z_a^{22}\equiv \left.{\rm Res}D^{22}\right|_{s=m_a^{*2}}
=\frac{\zeta(m_a^{*2})(m_a^{*2}-s_p)}{m_a^{*2}(X)-m_b^{*2}(X)}
\end{eqnarray}
where $\zeta(s)$ is assumed to be $\zeta(m_a^{*2}) \ne 0, \ne \infty$ (and
in fact it is the case).
Clearly the residue $|z_a^{22}|$ has a maximum value as a function of
$X$ when $m_a^*$ approaches nearest $m_b^*$.  {Such a phenomenon,
however, does not occur in a classical two-level problem. 
There only a level-repulsion takes place and residues do not exceed 1.
 The case
study for possible values of residues are given in
Appendix~\ref{sec:AppB}.}  

{At} the physical point $X=1$, the real part of $z^{22}$ is almost zero
and that of $z^{11}$ has a finite value as seen in
Fig.~\ref{fig:allz_vs_X}(a).
It seems that the physical
$\sigma$ state is purely composite 
and has no component of the elementary $\sigma$ meson.  
However, there is a finite component in the physical state or,
to be more precise, there must be a finite contribution from elementary
$\sigma$ to the amplitude, because Im\,$z^{22}$ is not zero as shown in
Fig.~\ref{fig:allz_vs_X}(b), and it {influences} the amplitude
{Eq.~(\ref{eq:T_full3})} as well as the real part {does}.
{Therefore} we show the modulus of the residues as functions
of the mixing parameter $X$ in Fig.~\ref{fig:allz_vs_X}(c) as a measure
{for} the contribution{s} from the basis states to the
amplitude.   For smaller $X$ the composite
$\sigma$ dominates the physical state ($|z^{11}|>|z^{22}|$), at $X\sim 0.5$ their
contributions become comparable ($|z^{11}|\sim|z^{22}|$), and at the
physical point $X=1$ the component of elementary $\sigma$ meson becomes
larger than that of the composite $\sigma$ meson ($|z^{11}|<|z^{22}|)$.  

In Figs.~\ref{fig:allz_vs_X}(d) and (e), we show the
square-root of the residues $\sqrt{z^{ii}}$ as functions of the mixing
parameter $X$ because it appears in the form of square-root in
Eq.~(\ref{eq:T_full3}).   
We
find that the strength of the imaginary parts of the 
square-root of the residues, Im$\sqrt{z^{11}}$ and Im$\sqrt{z^{22}}$,
are quite similar for all $X$, which allows us to forget about the
imaginary part when {discussing dominant component.}
As for the real part, we can clearly see again that the strengths of
Re$\sqrt{z^{11}}$ and Re$\sqrt{z^{11}}$ become similar at $X\sim 0.5$,
and the contribution from the elementary $\sigma$ becomes larger than
that of the composite $\sigma$ at the physical point $X=1$.
{As a result, when the mass of the elementary $\sigma$ meson
$m_0=550$~MeV is employed, the elementary nature becomes predominant in
the physical $\sigma$ meson.  In a later section, we will study the
$m_0$ dependence of the results.}

\subsection{a fate of pole-b and the number of poles}
Now, let us go back to the discussion of properties of {pole-b}.
\begin{figure}[htbp]
\includegraphics[width=0.85\linewidth]{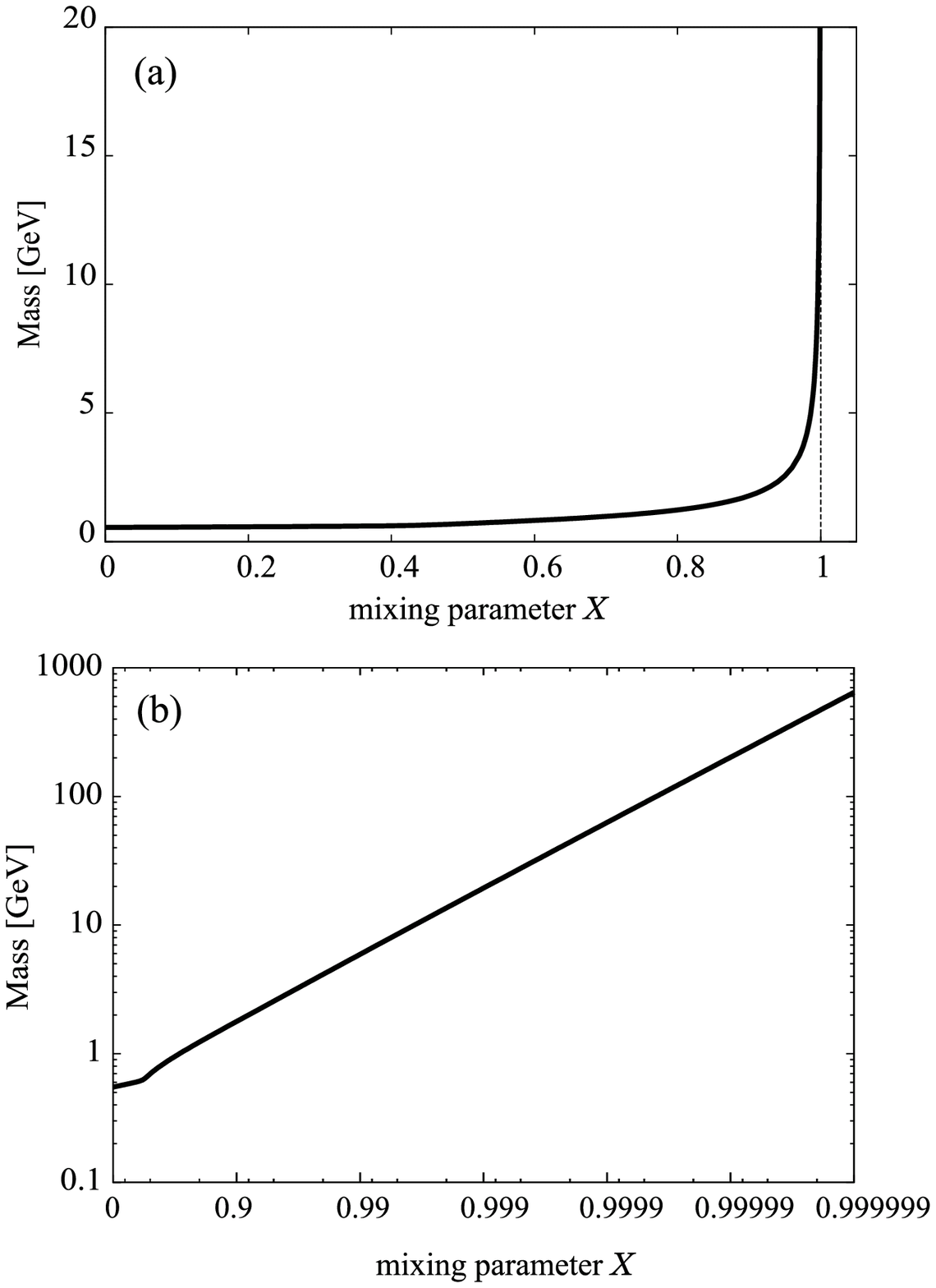}
\caption{Mass of pole-b as functions of mixing parameter $X$,
defined by the real part of the pole position in complex-energy plane,
 plotted (a) in linear plot and (b) in double logarithmic plot.
\label{fig:M_pole-b}}
\end{figure}
As shown in Figs.~\ref{fig:pole_flow_hyodo} and \ref{fig:pole_flow}, we
have two poles at finite $x$ and $X$, but one of these poles goes far
away from the energy region of interest as $x$ and/or $X$ is increased.
In Fig.~\ref{fig:M_pole-b}, as 
an example, we show the real part of the pole position of pole-b with
parameterization-(I\hspace{-.1em}I) (shown
in Fig.~\ref{fig:pole_flow}) as a function of the mixing parameter $X$.
{W}e find that pole-b goes {to} infinity in the limit $X=1$.
This {behavior} can be understood by {looking at the total}
potential {$v_{\rm con}+v_{\rm pole}$}.  
{For large $s$, where pole-b {is expected to} locate, 
the total potential is expanded {at $1/s \rightarrow 0$} as
\begin{eqnarray}
 v_{\rm con}+X v_{\rm pole} \xrightarrow[\text{large }s]{}& -& \frac{3(1-X)}{f_\pi^2} s \nonumber\\
&+&
\frac{1}{f_\pi^2}\left\{
m_\pi^2(1-6X)+m_0^2(3X+2)
\right\} \nonumber \\ &+& {\cal O}
\left(
\frac{1}{s}
\right). 
\label{eq:lim_v}
\end{eqnarray}
The first term is {the leading {and only} one for} the attraction, but it
disappears at $X=1$, and {so does} pole-b. }

{This can be seen in a better manner by using 
parameterization-(I).}
The first term of Eq.~(\ref{eq:A2}) yields the composite
$\sigma$ meson and the second term introduces the elementary $\sigma$
meson, which can be rewritten algebraically as
\begin{eqnarray}
 A(s)&=&-\frac{1}{f_\pi^2}(s-m_\pi^2) + x\frac{1}{f_\pi^2}
\frac{(s-m_\pi^2)^2}{s-m_0^2} \nonumber \\
&=&
 \frac{1}{f_\pi^2}\frac{(s-m_\pi^2)(m_0^2-m_\pi^2)}{s-m_0^2}
+(1-x)\frac{1}{f_\pi^2}
\frac{(s-m_\pi^2)^2}{s-m_0^2}  . \nonumber  \\
\label{eq:A3}
\end{eqnarray}
We may interpret the second line as that there are two different
``seeds'' having the same mass $m_0$ with different coupling strengths.  
Obviously, the second seed vanishes at $x=1$, leaving only one seed.

This can be understood physically
if we consider that the function $A(s)$
with $x=1$ can be also expressed as
\begin{equation}
 A(s)=\frac{1}{f_\pi^2}(m_0^2-m_\pi^2)+\frac{1}{f_\pi^2}(m_0^2-m_\pi^2)^2
\frac{1}{s-m_0^2}.
\end{equation}
This form is identical to what one obtains in the linear representation of
the sigma model, where the first term expresses the repulsive four-pion
interaction giving no composite $\sigma$ state dynamically.
There, the unitarized amplitude has the only one pole associated with the second term
corresponding to the 
elementary $\sigma$ meson acquiring a finite width {through the
coupling to the two-pion channel.} 
Because of the representation independence of the sigma
model~\cite{Donoghue:book}, we should have only one pole also in the
nonlinear representation.  
{In this case} the elementary $\sigma$ pole is considered to behave
like a counter term for the composite $\sigma$ pole without introducing
a second pole.

\begin{figure}[bp]
\includegraphics[width=0.95\linewidth]{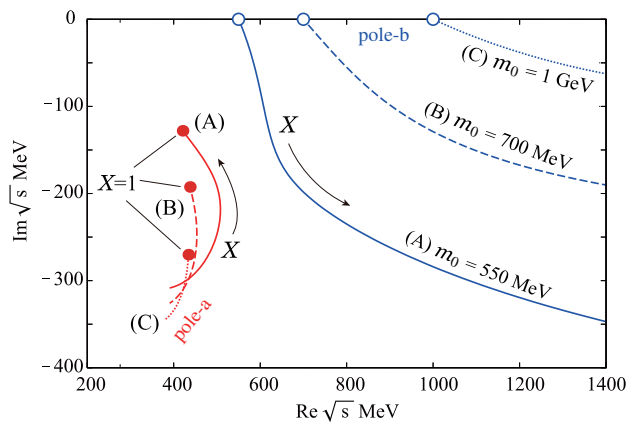}
\caption{(color online) Trajectories of the poles of $\hat{D}$ by
 changing the mixing parameter $X$ for different mass of the elementary
 $\sigma$
  (A)
 $m_0=550$~MeV (solid lines), (B) $m_0=700$~MeV (dashed
 lines), and (C) $m_0=1$~GeV (dotted lines). 
 The open circles indicate the mass of the
 elementary $\sigma$ meson, $m_0$, and  the solid circles indicate the
 pole position of the physical state at $X=1$ for each $m_0$ case.
\label{fig:pole_flow_ms}}
\end{figure}

\begin{figure*}[htbp]
\includegraphics[width=0.8\linewidth]{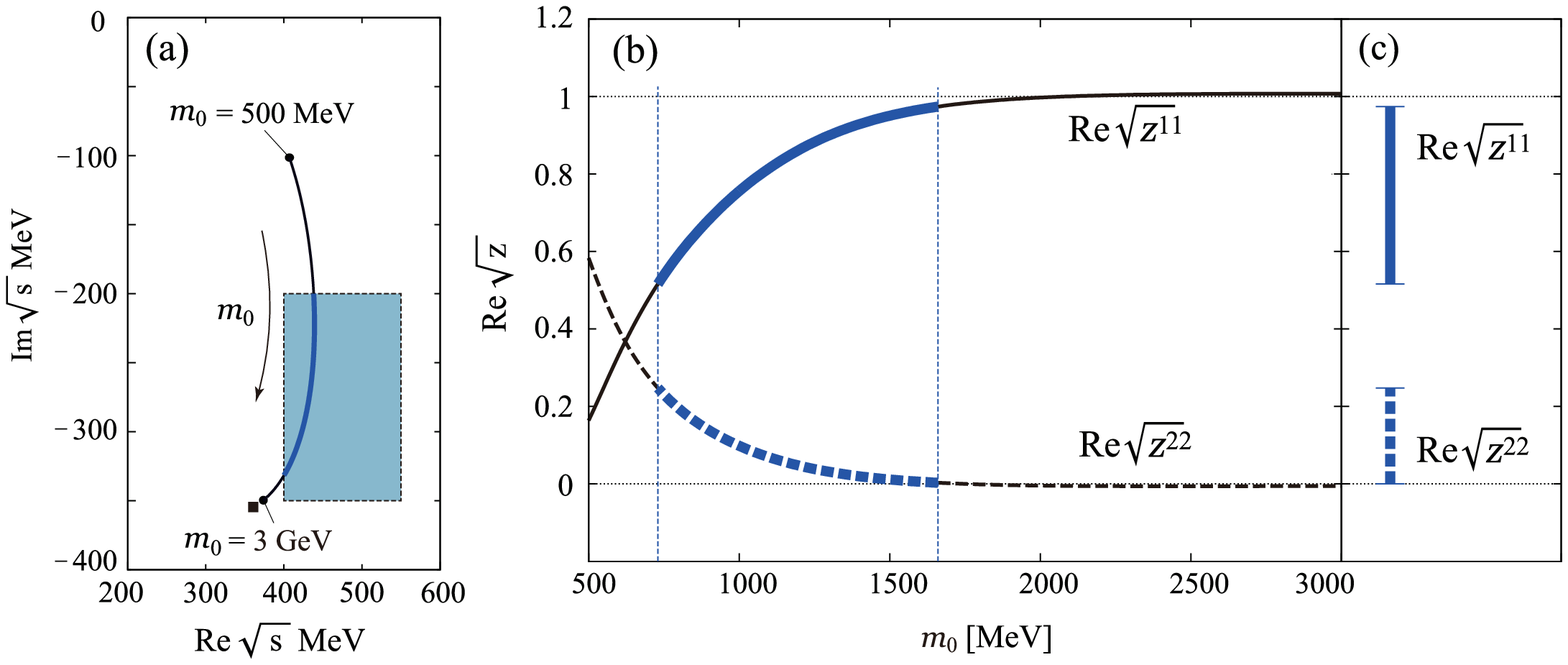}
\caption{(color online) (a) Trajectory of the poles of $\hat{D}$ at $X=1$ by changing
 the mass of the elementary $\sigma$, $m_0$.  The solid square indicates
 the pole position of the naive-composite $\sigma$ meson, which is also
 obtained in 
 $m_0\rightarrow\infty$ limit.  The (blue) hatched area denotes the
 energy range of the
 experimental value of the $\sigma$ mass $m_\sigma= 400$ -- 550 MeV and
 width $\Gamma_\sigma = 400$ -- 700 MeV in PDG.~\cite{PhysRevD.86.010001}. 
(b) The real parts of the square-root of the residues
 at the resulting pole at $X=1$ as functions of $m_0$.   
 The area between two vertical dashed-lines
 corresponds to the case in which the corresponding pole are within the
 errorbars of the PDG value~\cite{PhysRevD.86.010001}.
(c) The possible range of Re$\sqrt{z^{11}}$
 and  Re$\sqrt{z^{22}}$ when the resulting pole position is within errorbars
 of PDG value~\cite{PhysRevD.86.010001}.
\label{fig:vs_ms}}
\end{figure*}

Such a situation {occurs} only {at} $x=1$,
{implying that the number of poles}
depends on the coupling strength of $\sigma\pi\pi$.
{In fact}, we have found two poles in our previous study for the $a_1(1260)$ 
axial vector meson~\cite{Nagahiro:2011jn}. 
{There the interaction kernel for the $\pi\rho$ scattering is given
by} the Weinberg-Tomozawa interaction for $\pi\rho$ system and the $a_1$-pole
term\footnote{In Ref.~\cite{Nagahiro:2011jn}, we have
employed the value of the coupling strength of $a_1\pi\rho$ vertex
$g_{a_1\pi\rho}=0.26$~\cite{Sakai:2004cn,*Sakai:2005yt} instead of that
in the hidden Lagrangian~\cite{Bando:1987br}
$g_{a_1\pi\rho}=1/4$ {as in
Eq.~(\ref{eq:a1pole})}.}~\cite{Bando:1987br,Nagahiro:2011jn},  
\begin{gather}
 V_{WT} = - \frac{1}{4f_\pi^2}\left\{
3s - 2(m_\rho^2+m_\pi^2) - \frac{1}{s}(m_\rho^2-m_\pi^2)^2
\right\}, \\
V_{a_1 \text{-pole}} =
 \frac{1}{2f_\pi^2}(s-m_\rho^2)^2\frac{1}{s-m_{a_1}^2} .
\label{eq:a1pole}
\end{gather}
We {have found} two poles {for all $x$} even at $x=1$ where $x$
is introduced as 
$
 V_{\rm total} = V_{WT} + xV_{a_1 \text{-pole}}
$.  
Indeed, {the leading term of} the total potential $V_{\rm total}$ 
{remains} for large $s$ 
at $x=1$,
\begin{equation}
V_{WT} + V_{a_1 \text{-pole}} \xrightarrow[\text{large }s]{} - \frac{s}{4f_\pi^2} + 
\frac{m_{a_1}^2-m_\rho^2+m_\pi^2}{2f_\pi^2}
+{\cal
 O}\left(\frac{1}{s}\right).
\end{equation}
We find that, however, the pole-b in
Ref.~\cite{Nagahiro:2011jn} goes {to} infinity as well {at}
$x=3/2$, {where} the leading term of the total potential {vanishes,}
\begin{equation}
V_{WT} + \frac{3}{2}V_{a_1 \text{-pole}} \xrightarrow[\text{large }s]{}
 \frac{3m_{a_1}^2-4m_\rho^2+2m_\pi^2}{4f_\pi^2}+{\cal O}\left(\frac{1}{s}\right)
   \ \ .
\label{eq:a1_lim_v}
\end{equation}
So we conclude that, although we can define two basis states as
independent degrees of freedom, we do not necessarily have two resulting
states. 
Incidentally, it may be interesting to note that
the condition as Eqs.~(\ref{eq:lim_v}) or (\ref{eq:a1_lim_v})
is nothing but the unitarity condition for the tree-level amplitude.

\subsection{$m_0$  dependence}

\begin{figure*}[htbp]
\includegraphics[width=0.9\linewidth]{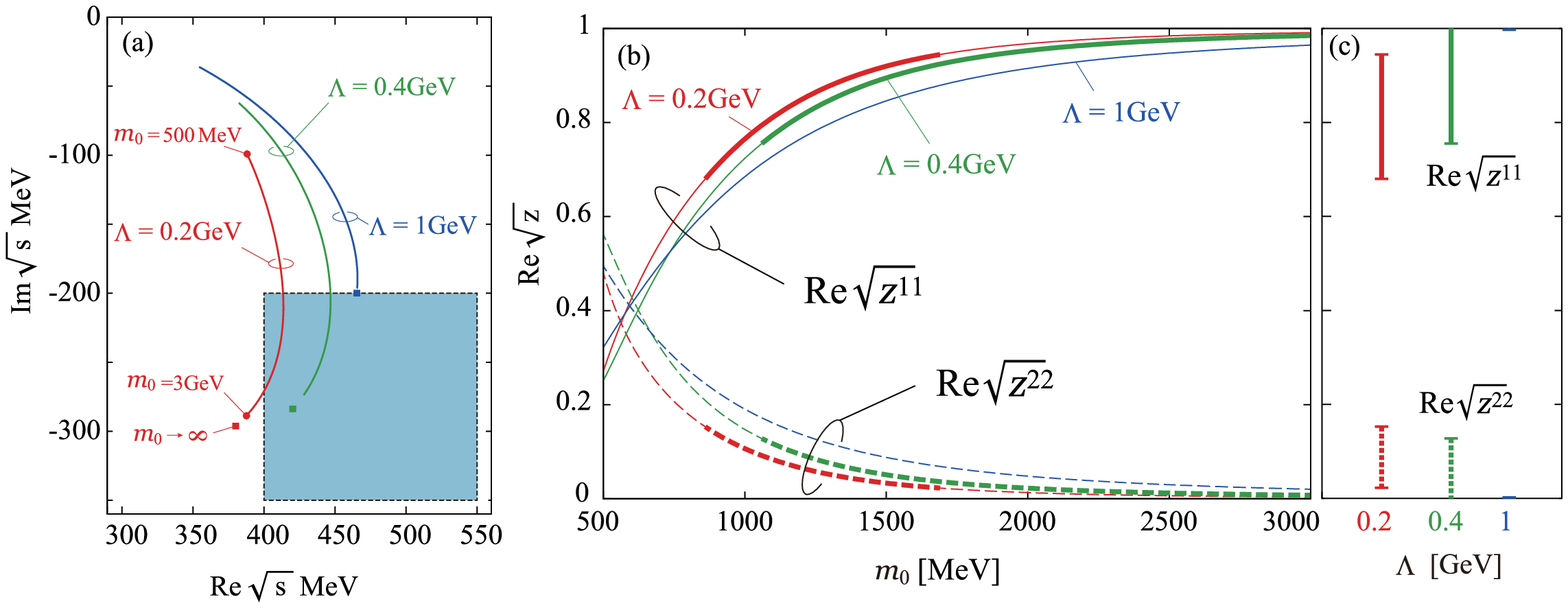}
\caption{(color online) (a) Trajectories of the poles of $\hat{D}$ at $X=1$
 by changing
 the mass of the elementary $\sigma$, $m_0$, for different cut-off
 values $\Lambda$.  The solid squares indicate the pole positions of the 
naive-composite $\sigma$ meson, which are also obtained in
 $m_0\rightarrow\infty$ limit.  The (blue) hatched area denotes the
 energy range of 
 experimental value of the $\sigma$ mass $m_\sigma=400$ -- 550~MeV and
 width $\Gamma_\sigma=400$ -- 700~MeV in PDG~\cite{PhysRevD.86.010001}. 
(b) The real parts of the square-root of the residues
 at the poles at $X=1$ as functions of $m_0$ for
 different $\Lambda$. 
 The thick lines corresponds to the case in which corresponding
 poles are within errorbars of the PDG value~\cite{PhysRevD.86.010001}.  (c)
 The possible range of Re$\sqrt{z^{11}}$ and 
 Re$\sqrt{z^{22}}$ for different $\Lambda$
 when the resulting pole position is within errorbars
 of the PDG value~\cite{PhysRevD.86.010001}.
\label{fig:vs_msL}}
\end{figure*}

{Next, we discuss the dependence of the resulting state on the mass
of the elementary $\sigma$, $m_0$.}
In Fig.~\ref{fig:pole_flow_ms}, we show the
pole-flow by changing the mixing parameter $X$ for different 
$m_0$ values, 550~MeV, 700~MeV, and 1~GeV.
As $m_0$ is increased, the range of the flow of pole-a becomes
narrower, and in the limit $m_0\rightarrow \infty$ the pole-a stays at a
single point while changing the parameter $X$.
In contrast, the flow of pole-b is getting further away from pole-a
and disappears at $X=1$ in any case.
{In Fig.~\ref{fig:vs_ms}(a),} we 
show the trajectories of the resulting pole at $X=1$ by changing $m_0$.
We can see that, as $m_0$ is increased, the {resulting} pole
{approaches} the naive-composite $\sigma$ pole (solid
square).   {The residues shown in Fig.}~\ref{fig:vs_ms}(b)
also {indicate} that the nature of the {physical} pole {approaches the}
naive-composite {one} in the heavy $m_0$ limit,
\begin{equation}
\sqrt{z^{11}}\xrightarrow[m_0 \rightarrow \infty]{} 1 \ ,
\sqrt{z^{22}}\xrightarrow[m_0 \rightarrow \infty]{} 0 \ .
\end{equation}
{In other words,} the wave function renormalization of the elementary
$\sigma$ becomes zero in the limit $m_0\rightarrow \infty$.
\cite{Weinberg:1965zz}.

In Fig.~\ref{fig:vs_ms}(a), we show the {region of the}
experimental values of the 
mass and width of the $\sigma$ meson shown in Particle Data Group
(PDG)~\cite{PhysRevD.86.010001}.  {W}e find
that, to reproduce the data  
within the present model, we need {a rather large mass of the} the
elementary $\sigma$ meson,
$m_0\sim 730$ -- 1660~MeV, {when} the natural {value of the}
subtraction constant {is used}, although 
the elementary component is always smaller than 
that of the composite one (${\rm Re}\sqrt{z^{11}}>{\rm
Re}\sqrt{z^{22}}$) as shown in Figs.~\ref{fig:vs_ms}(b) and (c).  The
values of residues are obtained as 
${\rm Re}\sqrt{z^{11}} \sim 0.52$ -- 0.98 and 
${\rm Re}\sqrt{z^{22}} \sim 0.25$ -- $2.0\times 10^{-4}$, respectively.

\subsection{cut-off dependence}
\begin{table*}[htbp]
\begin{center}
\begin{tabular}{c||c||c|c|c|c} 
$\Lambda$ & naive-composite pole & $m_0$ &  &  &  \\
${\rm [MeV]}$ & [MeV] & [MeV] & $\sqrt{z^{11}}$ & $\sqrt{z^{22}}$ & $|z^{22}|$ \\\hline\hline
200 & $380 -296i$ & 860 -- 1690 & $0.68+0.34i$ -- $0.94+0.16i$ &
		 $0.15-0.25i$ -- $0.022-0.091i$ & 0.085 -- $8.8\times 10^{-3}$ \\
300 & $400-293i$ & 930 $<$ & $0.71+0.32i$ & $0.15-0.24i $ & $7.8\times 10^{-2} <$ \\
400 & $420-284i$ & 1060$<$ & $0.76+0.29i$ & $0.13  -0.20i$ & $5.6\times 10^{-2} <$ \\
500 & $436-270i$ & $1220<$ & $0.80+0.24i$ & $0.10-0.16i$ & $3.7\times 10^{-2}<$ \\
600 & $448-255i$ & $1430<$ & $0.85+0.19i$ & $0.080-0.12i$ & $2.1\times 10^{-2}<$ \\
700 & $456-240i$ & $1710<$ & $ 0.90+0.15i$ & $0.058-0.088i$  & $1.1\times 10^{-2}<$ \\
800 & $461-226i$ & $2160<$ & $0.93+0.10i$ & $0.036-0.057i$ & $4.6\times 10^{-3}<$ \\
900 & $464-213i$ & $3060<$ & $0.97+0.05i$ & $0.018-0.029i$ & $1.2\times 10^{-3}<$ \\
1000 & $465-201i$ & $9240<$ & $ 0.997+0.0058i$  & $0.002-0.003i$  & $1.4\times 10^{-5}<$ \\
\end{tabular}
\end{center}
\caption{Critical values of $m_0$ for which the resulting pole is within
 the errorbars of the PDG value~\cite{PhysRevD.86.010001}
$m_\sigma=400$ -- $550$~MeV and  $\Gamma_\sigma=400$ -- 700~MeV for
 different $\Lambda$ in the third column.  The corresponding residues
 are also shown in the 
 following columns.  For 
 $\Lambda=200$~MeV case, there is an upper limit of $m_0$ as well as a
 lower limit, while there is only the lower limit for
 $\Lambda=300$~MeV -- 1~GeV. 
 In the second column, the naive-composite pole position (obtained
 with $m_0\rightarrow \infty$) is shown.
 If the naive-composite pole is within the errorbars of the PDG value,
 it means that there is no upper limit for $m_0$.
}\label{fig:residues}
\end{table*}

So far, we employed the natural value for the subtraction constant
$a(\mu)$ {in the} dimensional regularization
scheme~\cite{Hyodo:2008xr,Hyodo:2010jp}.  As 
{discussed} in Ref.~\cite{Hyodo:2008xr}, introducing the subtraction
constant different from the natural value is equivalent to the introduction
of the CDD pole term.
{Generally, the CDD pole can be expressed as an elementary particle
in a Lagrangian, which is regarded as a counter term of the
renormalization. The use of the different subtraction constant
(renormalization condition) is therefore absorbed into the mass of the
elementary particle. 
At this point, it should be emphasized that the resulting amplitude $t$
does not change.  However, by changing the subtraction constant and
accordingly the mass of the CDD pole (elementary particle), the mixing
ratio may change, since the wave function renormalization $z^{ii}$ depends on the
mass of the elementary particle as shown in the previous section.
In other
words, the theory alone cannot determine the mixing nature of the
resonance.  Hence we need an extra condition for them, in accordance
with which the mixing ratio is determined.  

In the present 
analysis, we have determined
the extra condition by employing the natural value for the subtraction
constant.  This corresponds to the use of the cut-off $\Lambda \sim
180$~MeV when the three-dimensional cut-off scheme is employed for the
loop function.  This value, however, seems somewhat too small as compared
to the typical hadronic scale $\sim 1$~GeV.  To make physical insight
clearer, in what follows we consider the three-dimensional cut-off
scheme with different values of $\Lambda$.
}

In Fig.~\ref{fig:vs_msL}(a), we show the trajectory of the resulting
poles by changing $m_0$ {when $\Lambda=200$~MeV is employed (red
line in the figure).
The flow is similar to that of the natural value case
shown in Fig.~\ref{fig:vs_ms}(a).
In Fig.~\ref{fig:vs_msL}(a) we also 
show the trajectories} for different cut-off
values {$\Lambda=400$~MeV and 1~GeV}.  We note that the
trajectories do not overlap despite the above discussion that the change
of the cut-off (regularization parameter) value {is} equivalent to the
change of {$m_0$}. 
One of the reason is that the three-dimensional 
cut-off used here breaks covariance of the theory.  
Then the difference of the loop function coming from the different
three-dimensional cut-off $\Lambda$ cannot be absorbed in the mass of
the elementary particle.  
Second reason, which is indeed the main source of the difference, is
found in the unitarization procedure employed in this work.  
Although the
equivalence holds only when summing up the diagrams including the
$s$-pole in the direction of $s$-channel, we sum up the $t$- and
$u$-exchange diagrams as well in the direction of $s$-channel.  
Therefore, the mass of the ``exchanged $\sigma$ meson'' cannot compensate
for the regularization parameter.
{But still} the difference coming
from these reasons is not {large} as shown in
Figs.~\ref{fig:vs_msL}(a) and (b).

Now we observe that
 the resulting pole tends to
approach the real axis as $\Lambda$ increases from 200~MeV to 1~GeV.
This behavior 
{resembles the flow of pole-a when $m_0$ is decreased.}
To put it the other way around, in the case of large cut-off
$\Lambda$
we need to have heavier mass of the elementary
$\sigma$ to reproduce the large width of the $\sigma$ meson of the 
PDG value~\cite{PhysRevD.86.010001}.
As summarized in  Table~\ref{fig:residues},
the component of the elementary $\sigma$ in the
resulting state becomes smaller for larger cut-off. 
%

It is interesting to see that, for cut-off values of hadronic scale, the
mass of the elementary $\sigma$ turns out to be rather heavy, and the
mixing ratio of the elementary component is quite small.
For instance, for $\Lambda=400$~MeV the mass of the elementary $\sigma$
is at least 1 GeV.
When $\Lambda=1$~GeV is employed, the mass of the elementary
$\sigma$ should be larger than 9~GeV and the resulting pole is
almost pure composite $\sigma$ meson, Re$\sqrt{z^{11}}\sim 1$ and
Re$\sqrt{z^{22}}\sim 0$.

\section{summary}
\label{sec:sum}
The mixing nature of the scalar resonance $\sigma$ {consisting of the 
$\pi\pi$ composite $\sigma$ and elementary $\sigma$} has been
investigated within the sigma model in the nonlinear representation.
We have
shown that the unitarized scattering amplitude can be expressed in the
form of the two-level problem. 
The mixing strengths of the composite and elementary
$\sigma$ mesons have been evaluated by means of the
residues of the full propagators of the composite and elementary
$\sigma$ mesons.  

{The major findings of this study are summarized as follows:
\begin{itemize}
 \item A rather heavy mass of the elementary $\sigma$ meson at least
       $m_0 \gtrsim 1$~GeV is preferred,
       if experimental data from PDG is to be reproduced.

 \item The elementary $\sigma$ component is small and the $\pi\pi$
	composite state dominates the physical $\sigma$.
\end{itemize}}

We also have found the following interesting properties as the
two-level problem of composite and elementary particles:
\begin{itemize}
 \item Even if we have two basis states of the composite and elementary
       particles, it happens that only one state remains,
       depending on the strength of the interactions.

\item The sigma model represented in the nonlinear base is described by
      the two basis states, which 
      generates the only one $\sigma$ state.
\end{itemize}

%
Here we would like to emphasize that whether a physical particle is more
elementary or composite like depends on the model Lagrangian and on how
we define the two bases.  In the sigma model in the nonlinear
representation, the resulting physical $\sigma$ can be more composite
like than the elementary like.  
%
We will further discuss this issue
elsewhere~\cite{nagahiro-hosaka}.

\section*{Acknowledgement}
This work is supported by the Grant-in-Aid for Scientific Research on
Priority Areas titled ``Elucidation of New Hadrons with a Variety of
Flavors'' (No.~24105707 for H.~N.) and (E01: No.~21105006 for A.~H.).

\appendix

\section{the scattering amplitude in the form of the two-level problem}
\label{sec:Appendix-A}
Here we show the derivation of the scattering amplitude in the form of
Eq.~(\ref{eq:T_fullM}).  To this end, we first {define} a
function $\omega$ by 
\begin{equation}
 v_{\rm con}=g_R \omega g_R \ ,
\end{equation}
where $g_R$ is the vertex function introduced in Eq.~(\ref{eq:t_WT}).
{By expressing $v_{\rm pole}$ in Eq.~(\ref{eq:t_pole}) similarly as}
\begin{equation}
 v_{\rm pole} = gD_\sigma g \ ,
\end{equation}
we rewrite the
potential terms in a matrix form as,
\begin{eqnarray}
 v_{\rm con}+v_{\rm pole} &=& g_R\omega g_R + gD_\sigma g \nonumber \\
&=&  (g_R,g)
\begin{pmatrix} \omega &  0 \\ 0 & D_\sigma  \end{pmatrix}
\begin{pmatrix}  g_R \\ g \end{pmatrix}.
\end{eqnarray}
{Here $g$ is the coupling of $\sigma\pi\pi$ as $g^2=3(s-m_\pi^2)^2/f_\pi^2$ and
$D_\sigma$ the free propagator of the elementary $\sigma$ meson
$D_\sigma^{-1}=s-m_0^2$}.
{T}hen the full amplitude {can be rewritten as}
\begin{widetext}
\begin{eqnarray}
 t &=& (v_{\rm con}+v_{\rm pole})
+ (v_{\rm con}+v_{\rm pole})G(v_{\rm con}+v_{\rm pole})
+ \cdots \nonumber \\ 
&=&
(g_R,  g )
\begin{pmatrix} \omega & 0 \\ 0  & D_\sigma \end{pmatrix}
\begin{pmatrix}  g_R \\ g \end{pmatrix} 
+ 
(g_R,g)
\begin{pmatrix} \omega & 0 \\ 0 & D_\sigma \end{pmatrix}
\begin{pmatrix}  g_R \\ g \end{pmatrix}
G
(g_R,g)
\begin{pmatrix} \omega & 0 \\ 0 & D_\sigma \end{pmatrix}
\begin{pmatrix}  g_R \\ g \end{pmatrix}
+\cdots \nonumber \\
&=& 
(g_R,g)
\left[
\begin{pmatrix} \omega & 0 \\ 0 & D_\sigma \end{pmatrix}
+
\begin{pmatrix} \omega & 0 \\ 0 & D_\sigma \end{pmatrix}
\begin{pmatrix} g_R G g_R & g_R G g \\ gGg_R & gGg\end{pmatrix}
\begin{pmatrix} \omega & 0 \\ 0 & D_\sigma \end{pmatrix}
+\cdots
\right]
\begin{pmatrix}  g_R \\ g \end{pmatrix} \\
&=&
(g_R,g)
\frac{1}{
\begin{pmatrix} \omega & 0 \\ 0 & D_\sigma \end{pmatrix}^{-1}
-\begin{pmatrix} g_R G g_R & g_R G g \\ gGg_R & gGg\end{pmatrix}
}
\begin{pmatrix}  g_R \\ g \end{pmatrix} \\
&=&
(g_R,g)
\frac{1}{
\begin{pmatrix} \omega^{-1} - g_R G g_R & 0 \\ 0 & D_\sigma^{-1} \end{pmatrix}
-\begin{pmatrix} 0  & g_R G g \\ gGg_R &  gGg  \end{pmatrix}
}
\begin{pmatrix}  g_R \\ g \end{pmatrix} \\
&=&
(g_R,g)
\frac{1}{
\begin{pmatrix} s-s_p & 0 \\ 0 & s-m_0^2 \end{pmatrix}
-\begin{pmatrix}  0 & g_R G g \\ gGg_R &  gGg  \end{pmatrix}
}
\begin{pmatrix}  g_R \\ g \end{pmatrix} 
\ ,
\end{eqnarray} 
\end{widetext} 
where in the last line we use the relation
\begin{eqnarray}
 \omega^{-1} - g_R G g_R = g_R^2(v_{\rm con}^{-1}-G) &=& g_R^2 t_{\rm
  composite}^{-1} \nonumber \\
&=& s-s_p \ \ .
\end{eqnarray} 
Finally we obtain the full scattering amplitude in a matrix form
 as
\begin{equation}
 t=(g_R,g)
\frac{1}{\hat{D}_0^{-1}-\hat{\Sigma}}
\begin{pmatrix}
 g_R  \\ g
\end{pmatrix} \ ,
\end{equation}
where 
\begin{equation}
 \hat{D}_0^{-1} = 
\begin{pmatrix} s-s_p & 0  \\ 0  & s-m_0^2 \end{pmatrix},\ 
 \hat{\Sigma} = 
\begin{pmatrix}
0 & g_R G g \\ 
g G g_R & g G g
\end{pmatrix} \ .
\end{equation}

\section{possible values of residues}
\label{sec:AppB}
{As discussed in Sec.~\ref{sec:results}, we have observed that the wave
function renormalization constants $z^{ii}$ take a maximum value at a
finite mixing parameter $X$.  Such a phenomenon does not occur in a
classical two-level problem.
In this appendix, we discuss the possible values for the wave
function renormalization in the two-level problem.}

{Let us consider a simple form of two-level Hamiltonian,
\begin{equation}
 \hat{H}=\begin{pmatrix} s-m_1^2 & 0 \\ 0 & s-m_2^2	\end{pmatrix}
-
\begin{pmatrix}0 & v \\ v& 0\end{pmatrix}
\end{equation}
where $m_1$ and $m_2$ denote the masses of two basis states and $v$
a mixing potential between them.
When $m_1$, $m_2$ and $v$ are real and energy-independent,
the mass difference between the two resulting
states becomes larger which is so-called the level-repulsion.
In this case, the wave function
renormalizations $z^{ii}$ for two states should take a value}
\begin{equation}
0<z^{11}(z^{22})<1 \ ,
\end{equation}
and satisfy the sum rule
\begin{equation}
 z^{11}+z^{22}=1 \ .
\end{equation}
When the mixing potential {is} complex
value {but energy-independent}, the sum rule is {still}
satisfied as 
\begin{equation}
{\rm Re}\,z^{11}+{\rm Re}\,z^{22}=1, \ \  {\rm Im}\,z^{11}+{\rm Im}\,z^{22}=0.
\end{equation}
{This sum rule is held} as far as the
potential is energy-independent.  In such a case two levels can
come closer and Re\,$z^{11}$ and/or
Re\,$z^{22}$ can take a value larger than 1 or {even becomes} negative.

In the present case {of the $\sigma$ system}, the sum rule is also
broken as
\begin{equation}
  z^{11}+z^{22} \ne 1 \ .
\end{equation}
This is due to the energy-dependence of the mixing potential {(self-energy)}
$\hat{\Sigma}$ in Eq.~(\ref{eq:D0&Sigma}). 
{One of origins of the energy dependence comes from the fact that
the self-energy $\hat{\Sigma}$ includes
the effect of {the} coupling to the two pion continuum.  By eliminating
the latter, $\hat{\Sigma}$ becomes energy dependent.}
This situation is the same that the wave function renormalization of
one-particle propagator {takes a different value than 1}
due to one-loop correction{s}.  {Another} source
is the
energy-dependence of the couplings $g(s)$ and $g_R(s)$. 
{In fact, this energy dependence strongly influences the value of
$z^{ii}$ when the poles flow over the wide energy range in
complex-energy plane.} 

\bibliography{a1_nature.bib}

\begin{thebibliography}{24}%
\makeatletter
\providecommand \@ifxundefined [1]{%
 \@ifx{#1\undefined}
}%
\providecommand \@ifnum [1]{%
 \ifnum #1\expandafter \@firstoftwo
 \else \expandafter \@secondoftwo
 \fi
}%
\providecommand \@ifx [1]{%
 \ifx #1\expandafter \@firstoftwo
 \else \expandafter \@secondoftwo
 \fi
}%
\providecommand \natexlab [1]{#1}%
\providecommand \enquote  [1]{``#1''}%
\providecommand \bibnamefont  [1]{#1}%
\providecommand \bibfnamefont [1]{#1}%
\providecommand \citenamefont [1]{#1}%
\providecommand \href@noop [0]{\@secondoftwo}%
\providecommand \href [0]{\begingroup \@sanitize@url \@href}%
\providecommand \@href[1]{\@@startlink{#1}\@@href}%
\providecommand \@@href[1]{\endgroup#1\@@endlink}%
\providecommand \@sanitize@url [0]{\catcode `\\12\catcode `\$12\catcode
  `\&12\catcode `\#12\catcode `\^12\catcode `\_12\catcode `\%12\relax}%
\providecommand \@@startlink[1]{}%
\providecommand \@@endlink[0]{}%
\providecommand \url  [0]{\begingroup\@sanitize@url \@url }%
\providecommand \@url [1]{\endgroup\@href {#1}{\urlprefix }}%
\providecommand \urlprefix  [0]{URL }%
\providecommand \Eprint [0]{\href }%
\providecommand \doibase [0]{http://dx.doi.org/}%
\providecommand \selectlanguage [0]{\@gobble}%
\providecommand \bibinfo  [0]{\@secondoftwo}%
\providecommand \bibfield  [0]{\@secondoftwo}%
\providecommand \translation [1]{[#1]}%
\providecommand \BibitemOpen [0]{}%
\providecommand \bibitemStop [0]{}%
\providecommand \bibitemNoStop [0]{.\EOS\space}%
\providecommand \EOS [0]{\spacefactor3000\relax}%
\providecommand \BibitemShut  [1]{\csname bibitem#1\endcsname}%
\let\auto@bib@innerbib\@empty
\bibitem [{\citenamefont {Beringer}\ \emph {et~al.}(2012)\citenamefont
  {Beringer} \emph {et~al.}}]{PhysRevD.86.010001}%
  \BibitemOpen
  \bibfield  {author} {\bibinfo {author} {\bibfnamefont {J.}~\bibnamefont
  {Beringer}} \emph {et~al.} (\bibinfo {collaboration} {Particle Data Group}),\
  }\href {\doibase 10.1103/PhysRevD.86.010001} {\bibfield  {journal} {\bibinfo
  {journal} {Phys. Rev. D}\ }\textbf {\bibinfo {volume} {86}},\ \bibinfo
  {pages} {010001} (\bibinfo {year} {2012})}\BibitemShut {NoStop}%
\bibitem [{\citenamefont {Basdevant}\ and\ \citenamefont
  {Lee}(1970)}]{PhysRevD.2.1680}%
  \BibitemOpen
  \bibfield  {author} {\bibinfo {author} {\bibfnamefont {J.~L.}\ \bibnamefont
  {Basdevant}}\ and\ \bibinfo {author} {\bibfnamefont {B.~W.}\ \bibnamefont
  {Lee}},\ }\href {\doibase 10.1103/PhysRevD.2.1680} {\bibfield  {journal}
  {\bibinfo  {journal} {Phys. Rev. D}\ }\textbf {\bibinfo {volume} {2}},\
  \bibinfo {pages} {1680} (\bibinfo {year} {1970})}\BibitemShut {NoStop}%
\bibitem [{\citenamefont {Achasov}\ and\ \citenamefont
  {Shestakov}(1994)}]{PhysRevD.49.5779}%
  \BibitemOpen
  \bibfield  {author} {\bibinfo {author} {\bibfnamefont {N.~N.}\ \bibnamefont
  {Achasov}}\ and\ \bibinfo {author} {\bibfnamefont {G.~N.}\ \bibnamefont
  {Shestakov}},\ }\href {\doibase 10.1103/PhysRevD.49.5779} {\bibfield
  {journal} {\bibinfo  {journal} {Phys. Rev. D}\ }\textbf {\bibinfo {volume}
  {49}},\ \bibinfo {pages} {5779} (\bibinfo {year} {1994})}\BibitemShut
  {NoStop}%
\bibitem [{\citenamefont {T\"ornqvist}\ and\ \citenamefont
  {Roos}(1996)}]{PhysRevLett.76.1575}%
  \BibitemOpen
  \bibfield  {author} {\bibinfo {author} {\bibfnamefont {N.~A.}\ \bibnamefont
  {T\"ornqvist}}\ and\ \bibinfo {author} {\bibfnamefont {M.}~\bibnamefont
  {Roos}},\ }\href {\doibase 10.1103/PhysRevLett.76.1575} {\bibfield  {journal}
  {\bibinfo  {journal} {Phys. Rev. Lett.}\ }\textbf {\bibinfo {volume} {76}},\
  \bibinfo {pages} {1575} (\bibinfo {year} {1996})}\BibitemShut {NoStop}%
\bibitem [{\citenamefont {Harada}\ \emph {et~al.}(1996)\citenamefont {Harada},
  \citenamefont {Sannino},\ and\ \citenamefont {Schechter}}]{PhysRevD.54.1991}%
  \BibitemOpen
  \bibfield  {author} {\bibinfo {author} {\bibfnamefont {M.}~\bibnamefont
  {Harada}}, \bibinfo {author} {\bibfnamefont {F.}~\bibnamefont {Sannino}}, \
  and\ \bibinfo {author} {\bibfnamefont {J.}~\bibnamefont {Schechter}},\ }\href
  {\doibase 10.1103/PhysRevD.54.1991} {\bibfield  {journal} {\bibinfo
  {journal} {Phys. Rev. D}\ }\textbf {\bibinfo {volume} {54}},\ \bibinfo
  {pages} {1991} (\bibinfo {year} {1996})}\BibitemShut {NoStop}%
\bibitem [{\citenamefont {Harada}\ \emph {et~al.}(1997)\citenamefont {Harada},
  \citenamefont {Sannino},\ and\ \citenamefont
  {Schechter}}]{PhysRevLett.78.1603}%
  \BibitemOpen
  \bibfield  {author} {\bibinfo {author} {\bibfnamefont {M.}~\bibnamefont
  {Harada}}, \bibinfo {author} {\bibfnamefont {F.}~\bibnamefont {Sannino}}, \
  and\ \bibinfo {author} {\bibfnamefont {J.}~\bibnamefont {Schechter}},\ }\href
  {\doibase 10.1103/PhysRevLett.78.1603} {\bibfield  {journal} {\bibinfo
  {journal} {Phys. Rev. Lett.}\ }\textbf {\bibinfo {volume} {78}},\ \bibinfo
  {pages} {1603} (\bibinfo {year} {1997})}\BibitemShut {NoStop}%
\bibitem [{\citenamefont {Oller}\ and\ \citenamefont
  {Oset}(1997)}]{Oller:1997ti}%
  \BibitemOpen
  \bibfield  {author} {\bibinfo {author} {\bibfnamefont {J.}~\bibnamefont
  {Oller}}\ and\ \bibinfo {author} {\bibfnamefont {E.}~\bibnamefont {Oset}},\
  }\href@noop {} {\bibfield  {journal} {\bibinfo  {journal} {Nucl.Phys.}\
  }\textbf {\bibinfo {volume} {A620}},\ \bibinfo {pages} {438} (\bibinfo {year}
  {1997})}\BibitemShut {NoStop}%
\bibitem [{\citenamefont {Igi}\ and\ \citenamefont
  {Hikasa}(1999)}]{PhysRevD.59.034005}%
  \BibitemOpen
  \bibfield  {author} {\bibinfo {author} {\bibfnamefont {K.}~\bibnamefont
  {Igi}}\ and\ \bibinfo {author} {\bibfnamefont {K.-i.}\ \bibnamefont
  {Hikasa}},\ }\href {\doibase 10.1103/PhysRevD.59.034005} {\bibfield
  {journal} {\bibinfo  {journal} {Phys. Rev. D}\ }\textbf {\bibinfo {volume}
  {59}},\ \bibinfo {pages} {034005} (\bibinfo {year} {1999})}\BibitemShut
  {NoStop}%
\bibitem [{\citenamefont {Oller}\ \emph {et~al.}(1999)\citenamefont {Oller},
  \citenamefont {Oset},\ and\ \citenamefont {Pelaez}}]{Oller:1998hw}%
  \BibitemOpen
  \bibfield  {author} {\bibinfo {author} {\bibfnamefont {J.~A.}\ \bibnamefont
  {Oller}}, \bibinfo {author} {\bibfnamefont {E.}~\bibnamefont {Oset}}, \ and\
  \bibinfo {author} {\bibfnamefont {J.~R.}\ \bibnamefont {Pelaez}},\ }\href
  {\doibase 10.1103/PhysRevD.59.074001} {\bibfield  {journal} {\bibinfo
  {journal} {Phys. Rev.}\ }\textbf {\bibinfo {volume} {D59}},\ \bibinfo {pages}
  {074001} (\bibinfo {year} {1999})}\BibitemShut {NoStop}%
\bibitem [{\citenamefont {Hyodo}\ \emph {et~al.}(2010)\citenamefont {Hyodo},
  \citenamefont {Jido},\ and\ \citenamefont {Kunihiro}}]{Hyodo:2010jp}%
  \BibitemOpen
  \bibfield  {author} {\bibinfo {author} {\bibfnamefont {T.}~\bibnamefont
  {Hyodo}}, \bibinfo {author} {\bibfnamefont {D.}~\bibnamefont {Jido}}, \ and\
  \bibinfo {author} {\bibfnamefont {T.}~\bibnamefont {Kunihiro}},\ }\href
  {\doibase 10.1016/j.nuclphysa.2010.09.016} {\bibfield  {journal} {\bibinfo
  {journal} {Nucl. Phys.}\ }\textbf {\bibinfo {volume} {A848}},\ \bibinfo
  {pages} {341} (\bibinfo {year} {2010})}\BibitemShut {NoStop}%
\bibitem [{\citenamefont {Nagahiro}\ \emph {et~al.}(2011)\citenamefont
  {Nagahiro}, \citenamefont {Nawa}, \citenamefont {Ozaki}, \citenamefont
  {Jido},\ and\ \citenamefont {Hosaka}}]{Nagahiro:2011jn}%
  \BibitemOpen
  \bibfield  {author} {\bibinfo {author} {\bibfnamefont {H.}~\bibnamefont
  {Nagahiro}}, \bibinfo {author} {\bibfnamefont {K.}~\bibnamefont {Nawa}},
  \bibinfo {author} {\bibfnamefont {S.}~\bibnamefont {Ozaki}}, \bibinfo
  {author} {\bibfnamefont {D.}~\bibnamefont {Jido}}, \ and\ \bibinfo {author}
  {\bibfnamefont {A.}~\bibnamefont {Hosaka}},\ }\href@noop {} {\bibfield
  {journal} {\bibinfo  {journal} {Phys.Rev.}\ }\textbf {\bibinfo {volume}
  {D83}},\ \bibinfo {pages} {111504} (\bibinfo {year} {2011})}\BibitemShut
  {NoStop}%
\bibitem [{\citenamefont {Bando}\ \emph {et~al.}(1988)\citenamefont {Bando},
  \citenamefont {Kugo},\ and\ \citenamefont {Yamawaki}}]{Bando:1987br}%
  \BibitemOpen
  \bibfield  {author} {\bibinfo {author} {\bibfnamefont {M.}~\bibnamefont
  {Bando}}, \bibinfo {author} {\bibfnamefont {T.}~\bibnamefont {Kugo}}, \ and\
  \bibinfo {author} {\bibfnamefont {K.}~\bibnamefont {Yamawaki}},\ }\href
  {\doibase 10.1016/0370-1573(88)90019-1} {\bibfield  {journal} {\bibinfo
  {journal} {Phys. Rept.}\ }\textbf {\bibinfo {volume} {164}},\ \bibinfo
  {pages} {217} (\bibinfo {year} {1988})}\BibitemShut {NoStop}%
\bibitem [{\citenamefont {Roca}\ \emph {et~al.}(2005)\citenamefont {Roca},
  \citenamefont {Oset},\ and\ \citenamefont {Singh}}]{Roca:2005nm}%
  \BibitemOpen
  \bibfield  {author} {\bibinfo {author} {\bibfnamefont {L.}~\bibnamefont
  {Roca}}, \bibinfo {author} {\bibfnamefont {E.}~\bibnamefont {Oset}}, \ and\
  \bibinfo {author} {\bibfnamefont {J.}~\bibnamefont {Singh}},\ }\href
  {\doibase 10.1103/PhysRevD.72.014002} {\bibfield  {journal} {\bibinfo
  {journal} {Phys. Rev.}\ }\textbf {\bibinfo {volume} {D72}},\ \bibinfo {pages}
  {014002} (\bibinfo {year} {2005})}\BibitemShut {NoStop}%
\bibitem [{\citenamefont {Nagahiro}\ \emph {et~al.}(2009)\citenamefont
  {Nagahiro}, \citenamefont {Roca}, \citenamefont {Hosaka},\ and\ \citenamefont
  {Oset}}]{Nagahiro:2008cv}%
  \BibitemOpen
  \bibfield  {author} {\bibinfo {author} {\bibfnamefont {H.}~\bibnamefont
  {Nagahiro}}, \bibinfo {author} {\bibfnamefont {L.}~\bibnamefont {Roca}},
  \bibinfo {author} {\bibfnamefont {A.}~\bibnamefont {Hosaka}}, \ and\ \bibinfo
  {author} {\bibfnamefont {E.}~\bibnamefont {Oset}},\ }\href {\doibase
  10.1103/PhysRevD.79.014015} {\bibfield  {journal} {\bibinfo  {journal} {Phys.
  Rev.}\ }\textbf {\bibinfo {volume} {D79}},\ \bibinfo {pages} {014015}
  (\bibinfo {year} {2009})}\BibitemShut {NoStop}%
\bibitem [{\citenamefont {Donoghue}\ \emph {et~al.}(1992)\citenamefont
  {Donoghue}, \citenamefont {Golowich},\ and\ \citenamefont
  {Holstein}}]{Donoghue:book}%
  \BibitemOpen
  \bibfield  {author} {\bibinfo {author} {\bibfnamefont {J.~F.}\ \bibnamefont
  {Donoghue}}, \bibinfo {author} {\bibfnamefont {E.}~\bibnamefont {Golowich}},
  \ and\ \bibinfo {author} {\bibfnamefont {B.~R.}\ \bibnamefont {Holstein}},\
  }\href@noop {} {\emph {\bibinfo {title} {Dynamics of the Standard Model}}}\
  (\bibinfo  {publisher} {Cambridge University Press, Cambridge},\ \bibinfo
  {year} {1992})\BibitemShut {NoStop}%
\bibitem [{\citenamefont {Hyodo}\ \emph {et~al.}(2008)\citenamefont {Hyodo},
  \citenamefont {Jido},\ and\ \citenamefont {Hosaka}}]{Hyodo:2008xr}%
  \BibitemOpen
  \bibfield  {author} {\bibinfo {author} {\bibfnamefont {T.}~\bibnamefont
  {Hyodo}}, \bibinfo {author} {\bibfnamefont {D.}~\bibnamefont {Jido}}, \ and\
  \bibinfo {author} {\bibfnamefont {A.}~\bibnamefont {Hosaka}},\ }\href@noop {}
  {\bibfield  {journal} {\bibinfo  {journal} {Phys. Rev.}\ }\textbf {\bibinfo
  {volume} {C78}},\ \bibinfo {pages} {025203} (\bibinfo {year}
  {2008})}\BibitemShut {NoStop}%
\bibitem [{\citenamefont {Weinberg}(1965)}]{Weinberg:1965zz}%
  \BibitemOpen
  \bibfield  {author} {\bibinfo {author} {\bibfnamefont {S.}~\bibnamefont
  {Weinberg}},\ }\href {\doibase 10.1103/PhysRev.137.B672} {\bibfield
  {journal} {\bibinfo  {journal} {Phys. Rev.}\ }\textbf {\bibinfo {volume}
  {137}},\ \bibinfo {pages} {B672} (\bibinfo {year} {1965})}\BibitemShut
  {NoStop}%
\bibitem [{\citenamefont {Lurie}\ and\ \citenamefont
  {Macfarlane}(1964)}]{Lurie:1964}%
  \BibitemOpen
  \bibfield  {author} {\bibinfo {author} {\bibfnamefont {D.}~\bibnamefont
  {Lurie}}\ and\ \bibinfo {author} {\bibfnamefont {A.~J.}\ \bibnamefont
  {Macfarlane}},\ }\href@noop {} {\bibfield  {journal} {\bibinfo  {journal}
  {Phys. Rev.}\ }\textbf {\bibinfo {volume} {136}},\ \bibinfo {pages} {B816}
  (\bibinfo {year} {1964})}\BibitemShut {NoStop}%
\bibitem [{\citenamefont {Lurie}(1968)}]{Lurie:book}%
  \BibitemOpen
  \bibfield  {author} {\bibinfo {author} {\bibfnamefont {D.}~\bibnamefont
  {Lurie}},\ }\href@noop {} {\emph {\bibinfo {title} {Particle and Fields}}}\
  (\bibinfo  {publisher} {Interscience Publishers, New York},\ \bibinfo {year}
  {1968})\BibitemShut {NoStop}%
\bibitem [{\citenamefont {Hyodo}\ \emph {et~al.}(2012)\citenamefont {Hyodo},
  \citenamefont {Jido},\ and\ \citenamefont {Hosaka}}]{PhysRevC.85.015201}%
  \BibitemOpen
  \bibfield  {author} {\bibinfo {author} {\bibfnamefont {T.}~\bibnamefont
  {Hyodo}}, \bibinfo {author} {\bibfnamefont {D.}~\bibnamefont {Jido}}, \ and\
  \bibinfo {author} {\bibfnamefont {A.}~\bibnamefont {Hosaka}},\ }\href
  {\doibase 10.1103/PhysRevC.85.015201} {\bibfield  {journal} {\bibinfo
  {journal} {Phys. Rev. C}\ }\textbf {\bibinfo {volume} {85}},\ \bibinfo
  {pages} {015201} (\bibinfo {year} {2012})}\BibitemShut {NoStop}%
\bibitem [{\citenamefont {Nagahiro}\ and\ \citenamefont
  {Hosaka}()}]{nagahiro-hosaka}%
  \BibitemOpen
  \bibfield  {author} {\bibinfo {author} {\bibfnamefont {H.}~\bibnamefont
  {Nagahiro}}\ and\ \bibinfo {author} {\bibfnamefont {A.}~\bibnamefont
  {Hosaka}},\ }\href@noop {} {}\bibinfo {note} {{in preparaion}}\BibitemShut
  {NoStop}%
\bibitem [{\citenamefont {Nawa}\ \emph {et~al.}(2011)\citenamefont {Nawa},
  \citenamefont {Ozaki}, \citenamefont {Nagahiro}, \citenamefont {Jido},\ and\
  \citenamefont {Hosaka}}]{Nawa:2011pz}%
  \BibitemOpen
  \bibfield  {author} {\bibinfo {author} {\bibfnamefont {K.}~\bibnamefont
  {Nawa}}, \bibinfo {author} {\bibfnamefont {S.}~\bibnamefont {Ozaki}},
  \bibinfo {author} {\bibfnamefont {H.}~\bibnamefont {Nagahiro}}, \bibinfo
  {author} {\bibfnamefont {D.}~\bibnamefont {Jido}}, \ and\ \bibinfo {author}
  {\bibfnamefont {A.}~\bibnamefont {Hosaka}},\ }\href@noop {} {\  (\bibinfo
  {year} {2011})},\ \Eprint {http://arxiv.org/abs/1109.0426} {arXiv:1109.0426
  [hep-ph]} \BibitemShut {NoStop}%
\bibitem [{\citenamefont {Sakai}\ and\ \citenamefont
  {Sugimoto}(2005{\natexlab{a}})}]{Sakai:2004cn}%
  \BibitemOpen
  \bibfield  {author} {\bibinfo {author} {\bibfnamefont {T.}~\bibnamefont
  {Sakai}}\ and\ \bibinfo {author} {\bibfnamefont {S.}~\bibnamefont
  {Sugimoto}},\ }\href {\doibase 10.1143/PTP.113.843} {\bibfield  {journal}
  {\bibinfo  {journal} {Prog. Theor. Phys.}\ }\textbf {\bibinfo {volume}
  {113}},\ \bibinfo {pages} {843} (\bibinfo {year}
  {2005}{\natexlab{a}})}\BibitemShut {NoStop}%
\bibitem [{\citenamefont {Sakai}\ and\ \citenamefont
  {Sugimoto}(2005{\natexlab{b}})}]{Sakai:2005yt}%
  \BibitemOpen
  \bibfield  {author} {\bibinfo {author} {\bibfnamefont {T.}~\bibnamefont
  {Sakai}}\ and\ \bibinfo {author} {\bibfnamefont {S.}~\bibnamefont
  {Sugimoto}},\ }\href {\doibase 10.1143/PTP.114.1083} {\bibfield  {journal}
  {\bibinfo  {journal} {Prog. Theor. Phys.}\ }\textbf {\bibinfo {volume}
  {114}},\ \bibinfo {pages} {1083} (\bibinfo {year}
  {2005}{\natexlab{b}})}\BibitemShut {NoStop}%
\end{thebibliography}%

\end{document}